\documentclass[twocolumn,amsmath,amssymb,floatfix,pra,showpacs,eqsecnum,footinbib,superscriptaddress]{revtex4}
\usepackage{amsmath,amssymb,natbib,bm,graphicx,url,psfrag}

\usepackage{color}

\newcounter{fig}

\newcommand{\RR}{\mathbb{R}}

\newcommand{\ZZ}{\mathbb{Z}}

\newcommand{\vc}[1]{\mathbf{#1}}

\newcommand{\abs}[1]{\left|#1\right|}
\newcommand{\bra}[1]{\left\langle \, #1 \,\right|}
\newcommand{\ket}[1]{\left|\, #1 \, \right\rangle}
\newcommand{\bket}[2]{\left\langle \, #1 \,|\, #2 \, \right\rangle}
\newcommand{\boket}[3]{\langle\, #1 \,|\, #2 \,|\, #3 \,\rangle}

\newcommand{\dg}{\dag}

\newcommand{\be}{\begin{equation}}
\newcommand{\ee}{\end{equation}}

\DeclareMathOperator{\bj}{J}
\DeclareMathOperator{\me}{me}

\DeclareMathOperator{\divg}{div}

\DeclareMathOperator{\rnd}{int}
\DeclareMathOperator{\mmod}{mod}
\DeclareMathOperator{\rre}{Re}
\DeclareMathOperator{\iim}{Im}

\begin{document}
\title{Charge insensitive qubit design derived from the Cooper pair box}
\author{Jens Koch} 
\affiliation{Departments of Physics and Applied Physics, Yale University, New Haven, Connecticut 06520, USA}
\author{Terri M.\ Yu}
\affiliation{Departments of Physics and Applied Physics, Yale University, New Haven, Connecticut 06520, USA}
\author{Jay Gambetta}
\affiliation{Departments of Physics and Applied Physics, Yale University, New Haven, Connecticut 06520, USA}
\author{A.\ A.\ Houck}
\affiliation{Departments of Physics and Applied Physics, Yale University, New Haven, Connecticut 06520, USA}
\author{D.\ I.\ Schuster}
\affiliation{Departments of Physics and Applied Physics, Yale University, New Haven, Connecticut 06520, USA}
\author{J.\ Majer}
\affiliation{Departments of Physics and Applied Physics, Yale University, New Haven, Connecticut 06520, USA}
\author{Alexandre Blais}
\affiliation{D\'epartement de Physique et Regroupement Qu\'eb\'ecois sur les Mat\'eriaux de Pointe,
Universit\'e de Sherbrooke, Sherbrooke, Qu\'ebec, Canada, J1K 2R1}
\author{M.\ H.\ Devoret}
\affiliation{Departments of Physics and Applied Physics, Yale University, New Haven, Connecticut 06520, USA}
\author{S.\ M.\ Girvin}
\affiliation{Departments of Physics and Applied Physics, Yale University, New Haven, Connecticut 06520, USA}
\author{R.\ J.\ Schoelkopf}
\affiliation{Departments of Physics and Applied Physics, Yale University, New Haven, Connecticut 06520, USA}
\date{September 26, 2007}
\begin{abstract}
Short dephasing times pose one of the main challenges in realizing a quantum computer. Different approaches have been devised to cure this problem for superconducting qubits, a prime example being the operation of such devices at optimal working points, so-called ``sweet spots.'' This latter approach led to significant improvement of $T_2$ times in Cooper pair box qubits [D.\ Vion et al., Science \textbf{296}, 886 (2002)]. Here, we introduce a new type of superconducting qubit called the ``transmon.'' Unlike the charge qubit, the transmon is designed to operate in a regime of significantly increased ratio of Josephson energy and charging energy $E_J/E_C$.  The transmon benefits from the fact that its charge dispersion decreases exponentially with $E_J/E_C$, while its loss in anharmonicity is described by a weak power law. As a result, we predict a drastic reduction in sensitivity to charge noise relative to the Cooper pair box and an increase in the qubit-photon coupling, while maintaining sufficient anharmonicity for selective qubit control. Our detailed analysis of the full system shows that this gain is not compromised by increased noise in other known channels.
\end{abstract}
\maketitle

\section{Introduction} 
Quantum information processing has emerged as a rich, exciting field due to both its potential applications in cryptography \cite{gisin} and computational speedup \cite{shor,deutsch,grover} and its value in designing quantum systems that can be used to study fundamental physics in previously inaccessible regimes of parameter space.  A promising physical paradigm for quantum computers is the superconducting Josephson junction qubit \cite{makhlin,devoret1,you}, which is classified into three types according to their relevant degree of freedom: charge \cite{bouchiat,nakamura}, flux \cite{friedman,wal}, and phase \cite{martinis}.  These systems have potentially excellent scalability thanks to well-established fabrication techniques such as photo and electron-beam lithography.  Unfortunately, superconducting qubits currently have coherence times which are not yet sufficient for error correction and scalable quantum computation.

There are several different strategies for enhancing the dephasing times in superconducting qubits. One approach \cite{martinis2} is to improve the properties of junctions and materials to eliminate excess sources of $1/f$ noise, whose origin remains unclear so far. This is a difficult and costly process, but it is likely to benefit a wide range of qubit designs when it is successful. A second approach is the elimination of linear noise sensitivity by operating qubits at optimal working points. So-called ``sweet-spot'' operation has already demonstrated \cite{vion3} an increase in dephasing times over previous experiments \cite{nakamura} which could be as large as three orders of magnitude, and illustrates that simple tailoring of quantum circuit design can boost qubit performance. In the long run, a combination of both strategies will probably be necessary to realize a scalable design for superconducting quantum computing.

In this paper, we follow the second approach and propose a new superconducting qubit: a transmission-line shunted plasma oscillation qubit,  which we call the \emph{transmon}.  In its design, it is closely related to the Cooper pair box (CPB) qubit in Ref. \cite{bouchiat}.  However, the transmon is operated at a significantly different ratio of Josephson energy to charging energy. This design choice, as we will show, should lead to dramatically improved dephasing times.

Two quantities crucial to the operation of a CPB are the anharmonicity and the charge dispersion of the energy levels.  A sufficiently large anharmonicity is needed to prevent qubit operations from exciting other transitions in the system.  The charge dispersion describes the variation of the energy levels with respect to environmental offset charge and gate voltage, and determines the sensitivity of the CPB to charge noise:  the smaller the charge dispersion, the less the qubit frequency will change in response to gate charge fluctuations. The magnitudes of charge dispersion and anharmonicity are both determined by the ratio of the Josephson energy to the charging energy $E_J/E_C$. Increasing this ratio reduces the (relative) energy level anharmonicity (which limits the speed of qubit operations).  However, it also decreases the overall charge dispersion and thus the sensitivity of the box to charge noise. This reduction is important, since even with operation at the first-order insensitive sweet spot, the Cooper-pair box can be limited by higher-order effects of the $1/f$ charge noise \cite{ithier}, and by the problem of quasiparticle poisoning, which can both shift the box from its optimal point.

The transmon exploits a remarkable fact:  the charge dispersion reduces \emph{exponentially} in $E_J/E_C$, while the anharmonicity only decreases \emph{algebraically} with a slow power law in $E_J/E_C$ \cite{cottet}.  Consequently, by operating the transmon at a much larger $E_J/E_C$ ratio than the CPB, one can greatly reduce charge noise sensitivity in the qubit while only sacrificing a small amount of anharmonicity. In fact, the charge dispersion can be so strongly suppressed that the qubit becomes practically insensitive to charge. This eliminates the need for individual electrostatic gates and tuning to a charge sweet spot, and avoids the susceptibility to quasiparticle poisoning, which both benefit the scaling to larger numbers of qubits. Amazingly, the transmon can at the same time increase the strength of electrical coupling between qubits, or between a qubit and a transmission line cavity serving as a bus.

Although the transmon has an $E_J/E_C$ ratio in between that of typical charge qubits and typical phase qubits, it is important to emphasize that the transmon is very different from both the CPB and phase qubits, including the capacitively shunted phase qubit proposed recently by Steffen et al.\ \cite{steffen}.  In the transmon, it is the natural anharmonicity of the cosine potential which allows qubit operations, whereas in the phase qubit, the $E_J/E_C$ ratio is so large that the required anharmonicity can only be restored by driving a current $I$ very close to $I_C$ through the system, creating a washboard potential, see Refs. \cite{makhlin}, \cite{devoret1} and \cite{you} for recent reviews.  The device presented in Ref.~\cite{steffen} operated at an energy ratio of $E_J/E_C \sim2\times10^4$, whereas the transmon will typically involve ratios of the order of several tens up to several hundreds and is operated without the need for any dc connections to the rest of the circuit. 
Thus, the transmon is a new type of superconducting qubit that should fix the main weakness of the CPB by featuring an exponential gain in the insensitivity to charge noise. The favorable insensitivity of CPBs to other noise sources such as critical current and flux noise is maintained (and further improved) in the transmon system, rendering it a very promising candidate for the next generation of qubits (see Table \ref{tab1}). 
A complementary proposal for using a capacitor to modify the $E_J/E_C$ ratio in superconducting flux qubits is put forward in Ref.~\cite{you2}.

The outline of the paper is as follows.  In Section \ref{sec:model}, we introduce the transmon and its effective quantum circuit. The solution of the corresponding Schr\"odinger equation and an analysis of its asymptotics enable a quantitative discussion of the charge dispersion and the anharmonicity in Sections \ref{sec:chdispersion} and \ref{sec:anharmonicity}, respectively. Section \ref{sec:flux} provides additional information about the flux degree of freedom in the split transmon, and the role of asymmetry in the two Josephson junctions. The circuit quantum electrodynamics (circuit-QED) physics \cite{blais1} of the transmon is investigated in Section \ref{sec:cqed}, where we show that despite the smallness of the charge dispersion, the transmon is expected to reach the strong-coupling limit of circuit QED. That is, we show that even though the transmon energy levels are insensitive to low frequency voltages, transitions between levels can strongly be driven by resonant radiation. We discuss in detail the modifications of the dispersive limit and the Purcell effect due to the increased $E_J/E_C$ ratio. The subsequent Sections \ref{sec:relaxation} and \ref{sec:dephasing} are devoted to the investigation of noise in the transmon system and its projected effect on relaxation ($T_1$) and dephasing ($T_2$) times. We conclude our paper with a summary and a comprehensive comparison of the transmon with existing superconducting qubits in Section \ref{sec:conclusions}.

\begin{figure}
    \centering
        \includegraphics[width=0.9\columnwidth]{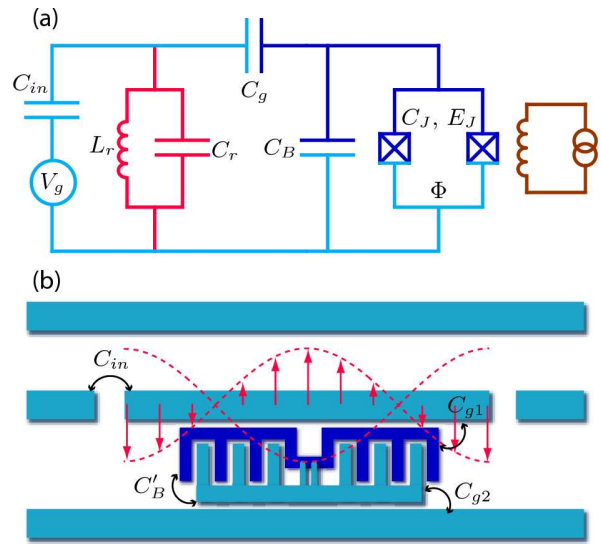}
    \caption{(Color online)  (a) Effective circuit diagram of the transmon qubit. The two Josephson junctions (with capacitance and Josephson energy $C_J$ and $E_J$) are shunted by an additional large capacitance $C_B$, matched by a comparably large gate capacitance $C_g$. (b) Simplified schematic of the transmon device design (not to scale), which consists of a traditional split Cooper-pair box, shunted by a short ($L\sim \lambda/20$) section of twin-lead transmission line, formed by extending the superconducting islands of the qubit. This short section of line can be well approximated as a lumped-element capacitor, leading to the increase in the capacitances $C_{g1}$, $C_{g2}$ and $C_B'$ and hence in the effective capacitances $C_B$ and $C_g$ in the circuit. \label{fig:circuit-diagram}}
\end{figure}

\section{From the Cooper pair box to the transmon}
\subsection{Model\label{sec:model}}
In close resemblance to the ordinary CPB (see e.g.\ Ref.\ \cite{devoret1}), the transmon consists of two superconducting islands coupled through two Josephson junctions, but isolated from the rest of the circuitry. This dc-SQUID setup allows for the tuning of the Josephson energy $E_J=E_{J,\text{max}}\abs{\cos (\pi\Phi/\Phi_0)}$ by means of an external magnetic flux $\Phi$. For simplicity, we initially assume that both junctions are identical. (The discussion of the general case including junction asymmetry is postponed until Section \ref{sec:flux}.)  Schematics of the device design and the effective quantum circuit for the transmon are depicted in Fig.~\ref{fig:circuit-diagram}.

As usual, the effective offset charge $n_g$ of the device, measured in units of the Cooper pair charge $2e$, is controlled by a gate electrode capacitively coupled to the island such that  $n_g=Q_r/2e+C_g V_g/2e$. Here $V_g$ and $C_g$ denote the gate voltage and capacitance, respectively, and $Q_r$ represents the environment-induced offset charge. 

The crucial modification distinguishing the transmon from the CPB is a shunting connection of the two superconductors via a large capacitance $C_B$, accompanied by a similar increase in the gate capacitance $C_g$. As shown in Appendix \ref{app:network}, the  effective Hamiltonian can be reduced to a form identical to that of the CPB system \cite{devoret2},
\be
\hat{H}=4E_C\left(\hat{n}-n_g\right)^2 - E_J \cos \hat{\varphi}.
\label{CPB-gen}
\ee
It describes the effective circuit of Fig.~\ref{fig:circuit-diagram}(a) in the absence of coupling to the transmission line (i.e.\ disregarding the resonator mode modeled by $L_r$ and $C_r$), and can be obtained from an analysis of the full network of cross capacitances as presented in Appendix \ref{app:network}.
The symbols $\hat{n}$ and $\hat{\varphi}$ denote the number of Cooper pairs transferred between the islands and the gauge-invariant phase difference between the superconductors, respectively.  By means of the additional capacitance $C_B$, the charging energy $E_C=e^2/2C_\Sigma$ ($C_\Sigma=C_J+C_B+C_g$) can be made small compared to the Josephson energy. In contrast to the CPB, the transmon is operated in the regime $E_J\gg E_C$. 

The qubit Hamiltonian, Eq.~\eqref{CPB-gen}, can be solved exactly in the phase basis in terms of Mathieu functions, see e.g.~Refs.~\cite{cottet,devoret1}.
The eigenenergies are given by
\be\label{energies}
E_m(n_g)=E_C\, a_{2[n_g+k(m,n_g)]}(-E_J/2E_C),
\ee
where $a_\nu(q)$ denotes Mathieu's characteristic value, and $k(m,n_g)$ is a function appropriately sorting the eigenvalues; see Appendix \ref{app:mathieu} for details. 
\begin{figure}[t]
    \centering
        \includegraphics[width=1.0\columnwidth]{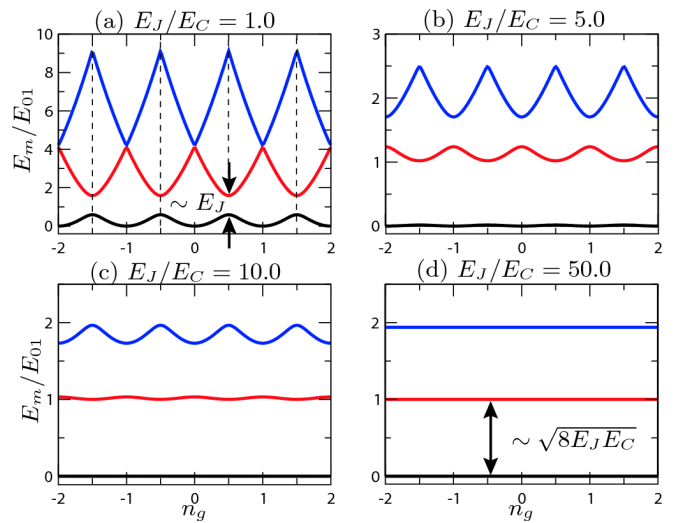}
    \caption{(Color online) Eigenenergies $E_m$ (first three levels, $m=0,1,2$) of the qubit Hamiltonian \eqref{CPB-gen} as a function of the effective offset charge $n_g$ for different ratios $E_J/E_C$. Energies are given in units of the transition energy $E_{01}$, evaluated at the degeneracy point $n_g=1/2$. The zero point of energy is chosen as the bottom of the $m=0$ level. The vertical dashed lines in (a) mark the charge sweet spots at half-integer $n_g$.}
    \label{fig:CPB-energies}
\end{figure} 
Plots for the lowest three energy levels $E_0$, $E_1$, and $E_2$, as a function of the effective offset charge $n_g$, are shown in Fig.~\ref{fig:CPB-energies} for several values of $E_J/E_C$. One clearly observes (i) that the level anharmonicity  depends on $E_J/E_C$, and (ii) that the total charge dispersion decreases very rapidly with $E_J/E_C$. Both factors (i) and (ii) influence the operation of the system as a qubit.  The charge dispersion immediately translates into the sensitivity of the system with respect to charge noise.  A sufficiently large anharmonicity is required for selective control of the transitions, and the effective separation of the Hilbert space into the relevant qubit part and the rest, $\mathcal{H}=\mathcal{H}_\text{q}\oplus\mathcal{H}_\text{rest}$.  In the following sections, we systematically investigate these two factors and show that there exists an optimal range of the ratio $E_J/E_C$ with sufficient anharmonicity and charge noise sensitivity drastically reduced when compared to the conventional CPB.

\subsection{The charge dispersion of the transmon\label{sec:chdispersion}}
The sensitivity of a qubit to noise can often be optimized by operating the system at specific points in parameter space. An example for this type of setup is the ``sweet spot'' exploited in CPBs \cite{vion2}.  In this case, the sensitivity to charge noise is reduced by biasing the system to the charge-degeneracy point $n_g=1/2$, see Fig.~\ref{fig:CPB-energies}(a).  Since the charge dispersion has no slope there, linear noise contributions cannot change the qubit transition frequency. With this procedure, the unfavorable sensitivity of CPBs to charge noise can be improved significantly, potentially raising $T_2$ times from the nanosecond to the microsecond range.  Unfortunately, the long-time stability of CPBs at the sweet spot still suffers from large fluctuations which drive the system out of the sweet spot and necessitate a resetting of the gate voltage.

Here, we show that an increase of the ratio $E_J/E_C$ leads to an \emph{exponential decrease} of the charge dispersion and thus a qubit transition frequency that is extremely stable with respect to charge noise; see Fig.~\ref{fig:CPB-energies}(d). In fact, with sufficiently large $E_J/E_C$, it is possible to perform experiments without any feedback mechanism locking the system to the charge degeneracy point.  In two recent experiments using transmon qubits, very good charge stability has been observed in the absence of gate tuning \cite{schuster,houck}. 

Away from the degeneracy point, charge noise yields first order corrections to the energy levels of the transmon and the sensitivity of the device to fluctuations of $n_g$ is directly related to the differential charge dispersion $\partial E_{ij}/\partial n_g$, as we will show in detail below. Here $E_{ij}\equiv E_j-E_i$ denotes the energy separation between the levels $i$ and $j$. As expected from a tight-binding treatment, the dispersion relation $E_m(n_g)$ is well approximated by a cosine in the limit of large $E_J/E_C$: 
\be\label{dispersion}
E_m(n_g)\simeq E_m(n_g=1/4) - \frac{\epsilon_m}{2} \cos (2\pi n_g),
\ee
where
\be
\epsilon_m\equiv E_m(n_g=1/2)-E_m(n_g=0)
\ee
gives the peak to peak value for the charge dispersion of the $m$th energy level.
To extract $\epsilon_m$, we start from the exact expression \eqref{energies} for the eigenenergies and study the limit of large Josephson energies. The asymptotics of the Mathieu characteristic values can be obtained by semi-classical (WKB) methods (see e.g.\ Refs.~\cite{goldstein,abramowitz1,connor}). The resulting charge dispersion is given by
\be\label{mresult}
\epsilon_m\simeq (-1)^m E_C \frac{2^{4m+5}}{m!}\sqrt{\frac{2}{\pi}}\left(\frac{E_J}{2E_C}\right)^{\frac{m}{2}+\frac{3}{4}}e^{-\sqrt{8E_J/E_C}},
\ee
valid for $E_J/E_C\gg1$. The crucial point of this result is the exponential decrease of the charge dispersion with $\sqrt{E_J/E_C}$. 

\begin{figure}[t]
    \centering
        \includegraphics[width=0.9\columnwidth]{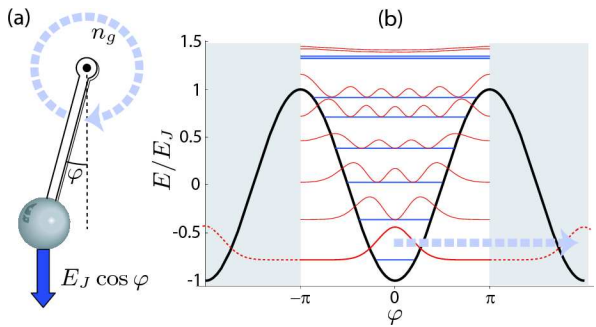}
    \caption{(Color online) (a) Rotor analogy for the transmon. The transmon Hamiltonian can be understood as a charged quantum rotor in a constant magnetic field $\sim n_g$. For large $E_J/E_C$, there is a significant ``gravitational'' pull on the pendulum and the system typically remains in the vicinity of $\varphi=0$. Only tunneling events between adjacent cosine wells (i.e.\ a full $2\pi$ rotor movement) will acquire an Aharonov-Bohm like phase due to $n_g$. The tunneling probability decreases exponentially with $E_J/E_C$, explaining the exponential decrease of the charge dispersion. (b) Cosine potential (black solid line) with corresponding eigenenergies and squared moduli of the eigenfunctions.
    \label{fig:rotor}}
\end{figure}

The physics behind this feature can be understood by mapping the transmon system to a charged quantum rotor, see Fig.~\ref{fig:rotor}. We consider a mass $m$ attached to a stiff, massless rod of length $l$, fixed to the coordinate origin by a frictionless pivot bearing. Using cylindrical coordinates $(r,\varphi,z)$, the motion of the mass is restricted to a circle in the $z=0$ plane with the polar angle $\varphi$ completely specifying its position. The rotor is subject to a strong homogeneous gravitational field $\vc{g}=g\vc{e}_x$ in $x$ direction, giving rise to a potential energy $V=-mgl\cos\varphi$. The kinetic energy of the rotor can be expressed in terms of its angular momentum along the $z$ axis, $\hat{L}_z=(\vc{r}\times\vc{p})\cdot\vc{e}_z = -i\hbar\frac{\partial}{\partial \varphi}$, so that the rotor's Hamiltonian reads
\be
H_\text{rot}=\frac{\hat{L}_z^2}{2ml^2} - mgl\cos\varphi.
\ee
Identifying the (integer-valued) number operator for Cooper pairs with the angular momentum of the rotor, $\hat{n}\leftrightarrow\hat{L}_z/\hbar$, and relating $E_J\leftrightarrow mgl$, $E_C\leftrightarrow(\hbar^2/8ml^2)$, one finds that the rotor Hamiltonian is identical to the transmon Hamiltonian with $n_g=0$. 

To capture the case of a nonzero offset charge, we imagine that the mass also carries an electrical charge $q$ and moves in a homogeneous magnetic field with strength $B_0$ in $z$ direction. Representing the magnetic field by the vector potential $\vc{A}=B_0(-y,x,0)/2$ (symmetric gauge) and noting that the vector potential enters the Hamiltonian according to 
\be
\vc{p}\to\vc{p}-q\vc{A}
\quad \Rightarrow \quad
L_z\to L_z +\frac{1}{2}qB_0l^2,
\ee 
one finds that the offset charge $n_g$ can be identified with $qB_0l^2/2\hbar$. This establishes a one-to-one mapping between the transmon system and the charged quantum rotor in a constant magnetic field. We emphasize that for the transmon (and CPB) the island charge is well-defined so that $\hat{n}$ has discrete eigenvalues and $\varphi$ is a compact variable leading to $\psi(\varphi)=\psi(\varphi+2\pi)$. In the rotor picture, this corresponds to the fact that the eigenvalues of the angular momentum $\hat{L}_z$ are discrete and that the ``positions" $\varphi$ and $\varphi+2\pi$ are identical. It is important to note that this mapping is different from the tilted washboard model used within the context of resistively shunted junctions, see e.g.~\cite{tinkham}, and must not be confused with this case. 

In the transmon regime, i.e.~large $E_J/E_C$, the dynamics of the rotor is dominated by the strong gravitational field.
Accordingly, small oscillation amplitudes around $\varphi=0$ are favored; see Fig.~\ref{fig:rotor}. Perturbation theory for small angles immediately leads to an anharmonic oscillator with quartic perturbation (Duffing oscillator).  [This method will be employed in Section \ref{sec:anharmonicity} to obtain the leading-order anharmonicity corrections.] However, the charge dispersion $\epsilon_m$ cannot be captured in such a perturbative picture. Within the perturbative approach (at any finite order) the $\varphi$ periodicity is lost and the angular variable becomes noncompact, $-\infty<\varphi<\infty$. Now, in the absence of the boundary condition $\psi(\varphi+2\pi)=\psi(\varphi)$ the vector potential can be eliminated by a gauge transformation. In other words, the effect of the offset charge $n_g$ only enters through the rare event of a full $2\pi$ rotation, in which case the system picks up an Aharonov-Bohm like phase. This corresponds to ``instanton" tunneling events through the cosine potential barrier to adjacent wells, and explains the WKB-type exponential decrease of the charge dispersion. It is interesting to note that the nonvanishing charge dispersion is truly a nonperturbative quantum effect, which can be ascribed to the discreteness of charge or equivalently to the peculiar role of the vector potential in quantum mechanics leading to the Aharonov-Bohm effect.

\begin{figure}[t]
    \centering
        \includegraphics[width=1.0\columnwidth]{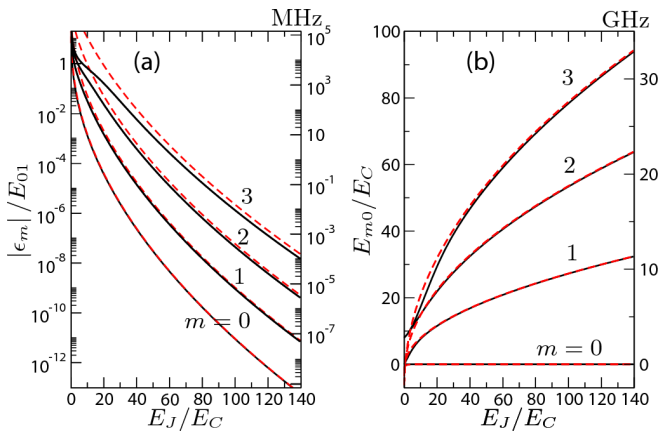}
    \caption{(Color online) Comparison of numerically exact and asymptotic expressions for the charge dispersion and energy levels. (a) Charge dispersion $\abs{\epsilon_m}$ as a function of the ratio $E_J/E_C$ for the lowest four levels. The solid curves depict the exact results using Mathieu characteristic values, the dashed curves represent the asymptotic expansion, Eq.~\eqref{mresult}. The right vertical scale gives the charge dispersion in MHz for a transition frequency of $7\,$GHz. (b) Energy level difference $E_{0m}=E_m-E_0$ at $n_g = 1/2$ as a function of the $E_J/E_C$ ratio. Solid curves show the exact results; dashed lines are based on the asymptotic expression \eqref{eigenenergiesas}. The vertical scale on the right-hand side gives the transition frequencies from the ground state to level $m$ in GHz, assuming a charging energy of $E_C/h=0.35\,\text{GHz}$. All numerical data are obtained for $n_g=1/2$.
    \label{fig:ripples}}
\end{figure}

The comparison between the exact result for the charge dispersion and the asymptotic expansion is depicted in Fig.~\ref{fig:ripples}(a). The requirements on the largeness of $E_J/E_C$ are seen to become stricter for increasing level index. For the transmon, we will mainly focus on the lowest two levels, for which Eq.~\eqref{mresult} constitutes a very good approximation when $E_J/E_C\ge 20$. Asymptotically, the differential charge dispersion $\partial E_{01}/\partial n_g$ is dominated by the contribution from the first excited level, so that from Eqs.~\eqref{dispersion} and \eqref{mresult} we have
\be\label{bgcurvature}
\frac{\partial E_{01}}{\partial n_g} \approx \pi\, \epsilon_1 \sin (2\pi n_g)
%-2^{9}E_C\sqrt{2\pi}\left(\frac{E_J}{2E_C}\right)^{5/4}e^{-\sqrt{8E_J/E_C}}\cos(2\pi n_g).
\ee
As a result, the maxima of $\abs{\partial E_{01}/\partial n_g}$ for $E_J/E_C=20,\,50,$ and $100$ are $7.3\times 10^{-2}E_C$, $1.5\times 10^{-4}E_C$, and $8.9\times 10^{-8}E_C$ respectively. These values should be contrasted with typical values of conventional CPBs operating in the limit of $E_J/E_C\alt1$ at the sweet spot $n_g=1/2$. In that case, the charge dispersion relation can be approximated by 
\be
E_{01}=\sqrt{[4E_C(2n_g-1)]^2 + E_J^2}.
\ee
At the sweet spot, the system is only sensitive to second-order noise, related to the curvature $\partial^2E_{01}/\partial n_g^2$ of the charge dispersion. This is given by $(8E_C)^2/E_J$, which for $E_J/E_C=1$ and $0.1$ leads to a curvature of $64E_C$ and $640E_C$, respectively. A comparison of these numbers demonstrates the remarkable robustness of the transmon to charge noise.  In Section \ref{sec:dephasing}, we will return to this point and translate our results into an estimate for the charge-noise induced dephasing time. As we will see, the transmon's dephasing time due to charge noise is exponentially increased in the parameter $E_J/E_C$. Another consequence of the drastically reduced charge sensitivity is that measurements of the island charge, see e.g.~\cite{schoelkopf,duty0}, cannot be employed to discriminate the qubit states. Even more, Eq.~\eqref{dispersion} implies that all higher derivatives of the eigenenergies with respect to offset charge become exponentially small. Thus, more general concepts such as the quantum capacitance  $C_q\sim\partial E_{i}^2/\partial n_g^2$ \cite{duty}, which works well at the CPB sweet spot, will fail. This simply reflects the fact that the inability of charge noise to ``measure" the qubit state and hence dephase it also means that a charge-based qubit readout becomes impossible.
Instead, we propose a dispersive readout via the cavity which we discuss in Section \ref{sec:dispersive}.

\subsection{Anharmonicity of the transmon\label{sec:anharmonicity}}
\begin{figure}[t]
    \centering
        \includegraphics[width=1.0\columnwidth]{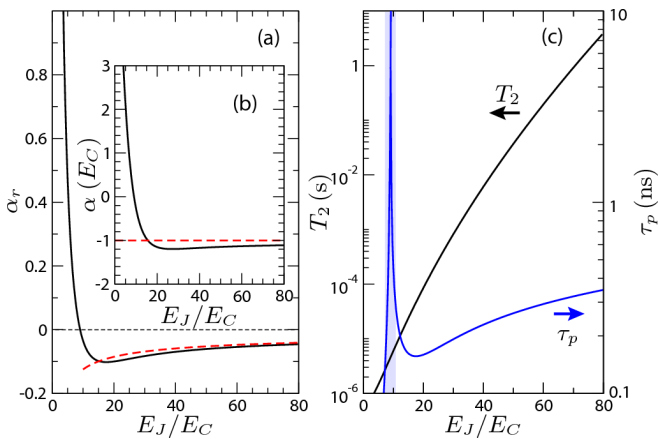}
    \caption{(Color online) Relative (a) and absolute (b) anharmonicity at the degeneracy point as a function of the ratio $E_J/E_C$. The solid curves show the exact results from Eq.~\eqref{energies}, the dashed curves depict the perturbative result from Eq.~\eqref{alpha-pert}. (c) Minimum pulse duration $\tau_p$ (blue/gray line) and dephasing time due to charge fluctuations $T_2$  (black line, Eq.~\eqref{charget2}) as a function of $E_J/E_C$. (Arrows indicate the corresponding axis.) Starting in the charge regime, an increase of the $E_J/E_C$ ratio initially leads to a strong increase in the minimum pulse duration, which diverges when the anharmonicity $\alpha$ crosses zero (``anharmonicity barrier''). Above the barrier, the operation time scales with a weak power law $\sim(E_J/E_C)^{1/2}$. At the same time, the inverse charge dispersion determining the dephasing time due to charge noise increases \emph{exponentially} in $(E_J/E_C)^{1/2}$.
    \label{fig:anharmonicity}}
\end{figure}

The impressive gain in charge-noise insensitivity by increasing $E_J/E_C$ has to be paid by a loss in anharmonicity. Sufficient anharmonicity is required to reduce the many-level system to a qubit, which ultimately sets a lower bound on the duration of control pulses. In the following we show that, in contrast to the charge dispersion, the anharmonicity only decreases with a weak power law. Therefore, we can find an $E_J/E_C$ range with significantly improved charge-noise insensitivity compared to the CPB as well as a sufficiently large anharmonicity.
We define the absolute and relative anharmonicity by 
\be\label{anharmonicity}
\alpha\equiv E_{12}-E_{01}, \qquad
\alpha_r\equiv\alpha/E_{01}.
\ee
Combining Eqs.~\eqref{energies} and \eqref{anharmonicity}, one concludes that the relative anharmonicity only depends on the effective offset charge and the energy ratio $E_J/E_C$. In the following, we investigate the anharmonicity evaluated at the charge-degeneracy point $n_g=1/2$, so that we can track the full crossover from the regular CPB regime (operating at the sweet spot) to the transmon regime \cite{footnote1}.
As shown in Fig.~\ref{fig:anharmonicity}(a), $\alpha_r$ then scales as $9(E_J/E_C)^{-1}$ in the small-$E_J/E_C$ limit. For $E_J/E_C\approx9$, it changes sign, indicating that for larger energy ratios the transition energy $E_{12}$ becomes smaller than $E_{01}$. The relative anharmonicity exhibits a shallow local minimum around $E_J/E_C\approx 17.5$ and asymptotically approaches zero for $E_J/E_C\to\infty$.

The scaling in this limit can be understood in terms of  perturbation theory in $(E_J/E_C)^{-1}\ll1$ \cite{footnote2}. Expanding the cosine in Eq.~\eqref{CPB-gen} around $\varphi=0$ up to fourth order, and treating the resulting quartic term in leading order perturbation theory, one obtains the following approximation for the eigenenergies [see Appendix\ \ref{app:perturb} for details]:
\be\label{eigenenergiesas}
E_m\simeq -E_J+\sqrt{8E_CE_J}\left(m+\frac{1}{2}\right) -\frac{E_C}{12}(6m^2+6m+3),
\ee
where $\omega_p=\sqrt{8E_CE_J}/\hbar$ is also known as the Josephson plasma frequency. A comparison of this approximation to the numerically exact result is shown in Fig.\ \ref{fig:ripples}(b).
The resulting asymptotic expressions for the absolute and relative anharmonicity are
\be\label{alpha-pert}
\alpha\simeq -E_C,\qquad
\alpha_r\simeq-(8E_J/E_C)^{-1/2},
\ee
depicted in Figs.\ \ref{fig:anharmonicity}(a) and (b) \cite{footnote3}.

With these relations and assuming that the transition frequency has a value of $\omega_{01}/2\pi\approx 10\,$GHz typical for experiments \cite{wallraff}, we can estimate the optimal $E_J/E_C$ range.  The resulting absolute anharmonicity is given by $\alpha = \hbar\omega_{01}\alpha_r$. From the frequency spread of a transform-limited pulse, we can estimate the corresponding minimum pulse duration to be $\tau_p\sim \abs{\omega_{01}\alpha_r}^{-1}$. For coherent control of the system, the pulse duration must remain small compared to $T_1$ and $T_2$. If the total dephasing times for the transmon were of the order of a few hundreds of nanoseconds as in recent experiments on CPBs \cite{wallraff}, reasonable pulse durations would be in the range of several tens of nanoseconds. It is interesting to note that significantly shorter microwave pulses are difficult to achieve, so that the large anharmonicity in CPBs cannot actually be exploited fully. Using a typical pulse length of $10\,\text{ns}$, we require a minimum anharmonicity of 
\be
\abs{\alpha_r^\text{min}}\sim (\tau_p \,\omega_{01})^{-1} \sim (10\,\text{ns}\times 2\pi\times10\,\text{GHz})^{-1} = 1/200\pi.
\ee
Employing Eq.~\eqref{alpha-pert}, we find that the energy ratio should satisfy $20 \alt E_J/E_C\ll 5\times10^4$, opening up a large range with exponentially decreased sensitivity to charge noise and yet sufficiently large anharmonicity for qubit operations. In other words, the transmon regime is reached without paying any serious penalty, and pulse generation techniques common for CPB qubits can directly be transferred to the transmon qubit. This is further illustrated in Fig.\ \ref{fig:anharmonicity}(c), where the inverse charge dispersion (determining $T_2$ due to charge noise, see Section \ref{sec:dephasing}) and the minimum pulse duration $\tau_p$ are plotted. As discussed in detail in Section \ref{sec:dephasing} below, dephasing times for the transmon are expected to be significantly larger as compared to CPBs. With the projected dephasing times of the order of $20\,\mu$s (most likely limited by critical current noise), pulse durations much longer then $10\,$ns could be used, making accessible even larger $E_J/E_C$ values and greater charge noise insensitivity.

 We emphasize that our considerations regarding $\tau_p$ provide a rough and simple order-of-magnitude estimate for the practical $E_J/E_C$ range. A more detailed analysis will also have to take into account exact pulse durations and shapes \cite{fazio,steffen2}. 

\subsection{Comparison: transmon and phase qubit\label{sec:numberphase}}
There exist some remarkable similarities between the transmon and typical phase qubits: both operate at $E_J/E_C\gg1$ (transmon: $\sim10^2$, phase qubit: $\sim10^4$ \cite{martinis,steffen}), so that their relevant energy scale is given to a good approximation by the plasma oscillation frequency, and both are well protected against charge noise. Given these similarities and based on the phase-number uncertainty \cite{louisell}, one might wonder whether the transmon is in fact more closely related to the phase qubit than to the CPB. We now show that this is not the case.

Structurally, the Hamiltonians of the CPB and the transmon are identical, see Eq.~\eqref{CPB-gen}. 
For both the CPB and the transmon the number operator counting the charge transferred across the junction is well-defined and the phase is compact, i.e.~the phase is restricted to the interval $0\le\varphi<2\pi$. Increasing the parameter $E_J/E_C$ smoothly maps the CPB into the transmon. By contrast, in the case of a phase qubit there is a dc connection between the two sides of the Josephson junction permitting a current (or equivalently flux) bias, and making the states with phases $\varphi$ and $\varphi+2\pi$ physically distinct. This topological difference makes it impossible to establish a continuous mapping between the transmon and the phase qubit via adiabatic changes of $E_J/E_C$. 

\begin{figure}[t]
    \centering
        \includegraphics[width=1.0\columnwidth]{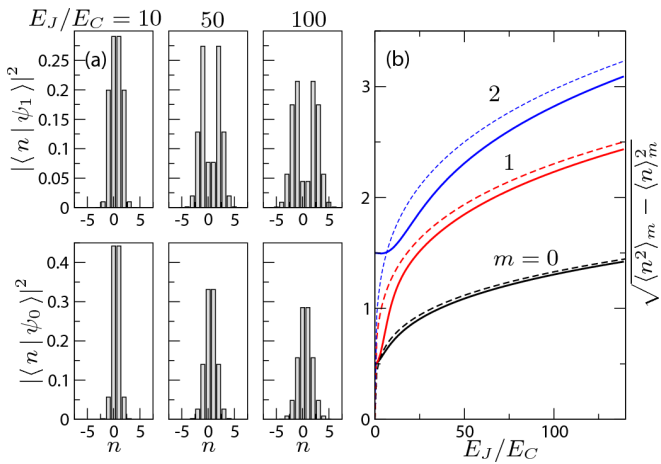}
    \caption{(Color online) Solutions to the qubit Hamiltonian \eqref{CPB-gen} in the charge basis. Panel (a) shows plots of the probabilities $\abs{\bket{n}{\psi_m}}^2$ for the presence of $n$ Cooper pairs when residing in the transmon eigenstates $m=0$ and $1$ for three different $E_J/E_C$ ratios ($n_g=1/2$). In the limit $E_J/E_C\gg1$, the solutions converge to (a discretized version of the) harmonic oscillator wavefunctions with increasing width. (b) Fluctuations of the number of Cooper pairs $n$ as a function of $E_J/E_C$ for the first three transmon levels. Solid lines show numerical exact results, dashed lines correspond to the asymptotic solution \eqref{chgfluc}. The Cooper pair number in the ground state (first excited state) fluctuates by approximately 1 (2) for an $E_J/E_C$ ratio of 100.
    \label{fig:chargesol}}
\end{figure}

We emphasize that the relationship between transmon and Cooper pair box does not imply that the eigenstates of the transmon are pure charge states. This is illustrated in Fig.~\ref{fig:chargesol}(a), where the overlap of the transmon eigenstates with pure charge states is shown. For increasing $E_J/E_C$, the transmon eigenstates spread over an increasing number of charge states. However, as derived in Appendix \ref{app:perturb} and depicted in Fig.~\ref{fig:chargesol}(b), the charge fluctuations only grow slowly as 
\be\label{chgfluc}
\sqrt{\langle \hat{n}^2 \rangle_m - \langle \hat{n} \rangle_m^2} \simeq \left(m+\frac{1}{2}\right)^{1/2}\left(\frac{E_J}{8E_C}\right)^{1/4},  
\ee
valid in the large $E_J/E_C$ limit.
As an example, for $E_J/E_C=100$ the number of Cooper pairs only fluctuates by approximately 1 and 2 in the ground and first excited state. In conclusion, the transmon is a CPB operated in the $E_J/E_C\gg1$ regime with charge fluctions of the order of unity.

\subsection{Split transmon: The flux degree of freedom and junction asymmetry\label{sec:flux}}
In the previous sections, we have ignored the fact that the proposed transmon design in fact involves \emph{two} Josephson junctions. Strictly speaking, this is only appropriate if the two junctions are identical (i.e.\ they feature the same Josephson coupling energy $E_{J1}=E_{J2}$). In that case the contributions simply add up and our previous treatment is valid. However, with current junction fabrication techniques junction parameters vary and typically lead to junction asymmetries up to $d\equiv\frac{E_{J2}-E_{J1}}{E_{J1}+E_{J2}}\simeq\pm10\%$. In the following, we discuss the effects of this asymmetry, which are also known in the context of CPBs; see e.g.\ \cite{vion3}.

The case of asymmetric junctions is described by replacing the cosine term in the Hamiltonian \eqref{CPB-gen} by the Josephson Hamiltonian
\be
\hat{H}_J=-E_{J1} \cos \hat{\phi}_1 -E_{J2} \cos \hat{\phi}_2,
\ee
where $\phi_{1,2}$ now describe the individual superconducting phase differences across the junctions 1 and 2; see e.g.\ \cite{tinkham}. The usual argument of flux quantization then leads to the condition 
\be
\phi_1-\phi_2 = 2\pi n + 2\pi\Phi/\Phi_0,
\ee
with integer $n$, and $\Phi$, $\Phi_0=h/2e$ denoting the magnetic flux through the SQUID-like ring and the superconducting flux quantum, respectively.  Defining the effective phase difference of the device as $\varphi=(\phi_1+\phi_2)/2$ and $E_{J\Sigma}=E_{J1}+E_{J2}$, the Josephson Hamiltonian may be rewritten as
\begin{align}
&\hat{H}_J=-E_{J\Sigma}\left[  \cos(\pi\Phi/\Phi_0) \cos \hat{\varphi} + d\, \sin (\pi\Phi/\Phi_0)  \sin \hat{\varphi}\right]\nonumber\\
&=-E_{J\Sigma}\cos \left(\frac{\pi\Phi}{\Phi_0}\right)\sqrt{1+d^2 \tan^2 \left(\frac{\pi\Phi}{\Phi_0}\right)}\cos(\hat{\varphi}-\varphi_0),
\label{josham}
\end{align}
where the phase $\varphi_0$ is determined by $\tan \varphi_0=d\tan (\pi\Phi/\Phi_0)$. For constant magnetic flux, this phase can be eliminated by a shift of variables. As a result, our previous results for the symmetric transmon ($d=0$) translate to the general case by substituting the Josephson energy by
\be
E_J \to E_{J\Sigma}\cos \left(\frac{\pi\Phi}{\Phi_0}\right)\sqrt{1+d^2 \tan^2 \left(\frac{\pi\Phi}{\Phi_0}\right)}.
\ee
Interestingly, for asymmetric junctions the flux dependence of $\varphi_0$ may allow for additional qubit control, not involving the resonator, by applying ac magnetic fields. As with all extra control channels, junction asymmetry leads to an additional qubit decay channel from flux fluctuations, which we will discuss in Section \ref{sec:relaxation}.

\section{Circuit QED for the transmon\label{sec:cqed}}
In close analogy to the situation of the CPB, embedding the transmon in a superconducting transmission line resonator opens up the possibility of control and readout of the qubit state -- a scenario that has been termed \emph{circuit QED} \cite{wallraff,blais1}. We start from the quantum-circuit Hamiltonian for a transmon attached to a superconducting transmission line, depicted in Fig.\ \ref{fig:circuit-diagram}(a). With the Josephson junctions centered in the transmission line, the relevant resonator mode is the $\ell=2$ mode (voltage antinode at the center of the resonator), and it can be described by a simple LC oscillator \cite{blais1}. In the realistic limit of large resonator capacitance $C_r\gg C_\Sigma$, the quantization of the circuit results in the effective Hamiltonian
\begin{align}\label{ham1}
\hat{H}=&4E_C (\hat{n}-n_g)^2 - E_J \cos \hat{\varphi}   +\hbar\omega_r\hat{a}^\dag \hat{a}\nonumber\\
&+ 2\beta eV^0_\text{rms} \hat{n}(\hat{a}+\hat{a}^\dag),
\end{align}
see Appendix \ref{app:network} for the detailed derivation. Here, $\omega_r=1/\sqrt{L_rC_r}$ denotes the resonator frequency, and $\hat{a}$ ($\hat{a}^\dag$) annihilates (creates) one photon in the transmission line. The root mean square voltage of the local oscillator is denoted by $V^0_\text{rms}=\sqrt{\hbar\omega_r/2C_r}$. The parameter $\beta$ is defined as the ratio of the gate capacitance and the total capacitance,  $\beta=C_g/C_\Sigma$. 

Rewriting the Hamiltonian in the basis of the uncoupled transmon states $\ket{i}$, one obtains the generalized Jaynes-Cummings Hamiltonian
\be\label{hamiltonian-before}
\hat{H}= \hbar\sum_j\omega_j \ket{j}\bra{j} +\hbar\omega_r \hat{a}^\dag \hat{a} 
 + \hbar\sum_{i,j}  g_{ij}\ket{i}\bra{j}(\hat{a} +\hat{a}^\dag),
\ee
with coupling energies
\be\label{couplings}
\hbar g_{ij}=2\beta eV^0_\text{rms}\boket{i}{\hat{n}}{j}=\hbar g_{ji}^*.
\ee
The general expression \eqref{hamiltonian-before} can be significantly simplified by examining the matrix elements $\boket{i}{\hat{n}}{j}$, and invoking the rotating wave approximation. First, note that the asymptotic behavior of the matrix elements  can be evaluated within the perturbative approach introduced in Section \ref{sec:anharmonicity} and detailed in Appendix \ref{app:perturb}. Asymptotically, the number operator assumes the form $\hat{n}=-i(E_J/8E_C)^{1/4}(\hat{b}-\hat{b}^\dag)/\sqrt{2}$, so that
\begin{align}\label{n10}
\abs{\boket{j+1}{\hat{n}}{j}}&\approx\sqrt{\frac{j+1}{2}}\left( \frac{E_J}{8E_C} \right)^{1/4},\\ \abs{\boket{j+k}{\hat{n}}{j}}&\xrightarrow{E_J/E_C\to\infty}0\label{n20}
\end{align}
with $\abs{k}>1$, and $\hat{b}$, $\hat{b}^\dag$ denoting the annihilation and creation operator for the harmonic oscillator approximating the transmon. It is interesting to note that off-diagonal matrix elements with an even difference $k$ between states fall off exponentially, which can be understood from the point of view of the parity of the states, as well as from the fact that the leading anharmonic perturbation $(\hat{b}+\hat{b}^\dag)^4$ does not mix even and odd states. By contrast, matrix elements with odd $k>1$ show a slower power-law type decay as $E_J/E_C\to\infty$. This is illustrated  in Fig.~\ref{fig:melements}(a). From Eqs.\ \eqref{n10} and \eqref{n20} we conclude that nearest-neighbor coupling $g_{i,i\pm1}$ constitutes the only relevant coupling in the large $E_J/E_C$ limit. 

Finally, employing the rotating wave approximation to eliminate terms describing the simultaneous excitation (de-excitation) of both the transmon and the resonator, we arrive at the effective generalized Jaynes-Cummings Hamiltonian
\begin{align}
\label{hamiltonian-before2}
\hat{H}=& \hbar\sum_j\omega_j \ket{j}\bra{j} +\hbar\omega_r \hat{a}^\dag \hat{a} \\\nonumber
 &+ \bigg[ \hbar\sum_{i} g_{i,i+1}\ket{i}\bra{i+1}\hat{a}^\dag +\text{h.c.}\bigg].
\end{align}
We remark that in contrast to the regular Jaynes-Cummings Hamiltonian, this generalized version does not allow for an exact analytical solution. The Hamiltonian remains block diagonal. However, due to the presence of many transmon levels and nearest neighbor coupling, the tridiagonal blocks grow in size, and the solution of the general case requires numerical methods. The dispersive limit allows for analytical solutions as we show in Section \ref{sec:dispersive}.

\subsection{The coupling strength of the transmon\label{sec:coupl}}
Despite the exponentially decreasing charge dispersion for large $E_J/E_C$, the coupling between cavity and transmon, expressed by the coupling energies $\hbar g_{ij}$, does \emph{not} become small but in fact even increases. This is a central message of this paper, and it is crucial for utilizing the transmon system as an actual qubit. 

Mathematically, the couplings $g_{ij}$ are determined by a prefactor containing the capacitance ratio $\beta$, the rms voltage of the local oscillator $V^0_\text{rms}$, and by a matrix element of the number operator for Cooper pairs, which depends on the energy ratio $E_J/E_C$. It is interesting to note that there is a fundamental upper bound to the magnitude of the prefactor. To see this, we rewrite
\be
\frac{2\beta eV^0_\text{rms}}{\hbar\omega_r} = 4\beta \sqrt{\frac{\alpha}{\epsilon_r}},
\ee
where we have used the relation $C_r=\epsilon_r\pi/2\omega_rZ_0$ valid for a half-wave transmission line resonator, and $\alpha=e^2/4\pi\epsilon_0\hbar c$ denotes the fine structure constant. For realistic values of the effective dielectric constant, this limits the relative coupling to about 10\% of the resonator frequency $\omega_r$.

\begin{figure}[t]
    \centering
        \includegraphics[width=1.0\columnwidth]{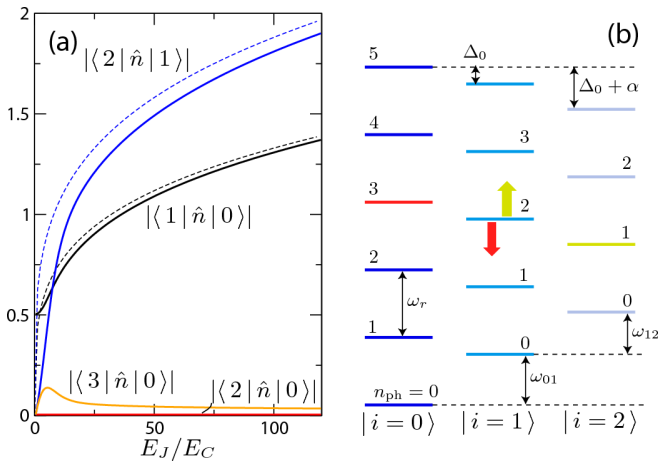}
    \caption{(Color online) (a) Off-diagonal matrix elements of the Cooper-pair number operator as a function of the energy ratio $E_J/E_C$ ($n_g=1/2$). Solid curves represent the exact result, dashed curves depict the asymptotic behavior, Eq.~\eqref{n10}. The results illustrate that coupling between neighboring transmon states is the only relevant coupling in the limit of large $E_J/E_C$.(b) Level scheme for the coupled transmon system. Transmon states are denoted by $\ket{i}$, $i=1,2,3$ for the ground, first and second excited state. Photon numbers $n_\text{ph}$ in the cavity are plotted vertically. The two arrows on the $\ket{i=1,n=2}$ level illustrate the perturbative level repulsions leading to the dispersive shift.
    \label{fig:melements}}
\end{figure}

While the magnitude of the prefactor is limited, the relevant matrix elements between neighboring transmon states exhibit an approximate power-law increase as a function of $E_J/E_C$, as depicted in Fig.~\ref{fig:melements}. 
As a result of Eq.~\eqref{n10}, the couplings $g_{i,i+1}$ asymptotically increase as $(E_J/E_C)^{1/4}$ \cite{footnotex}. Comparing this  with Eq.~\eqref{chgfluc}, we find that the increase of $g_{i,i+1}$ can be directly related to the increase of the charge number fluctuations. We emphasize that the parameters $\beta$ and $E_J/E_C$ can be tuned \emph{separately}; $\beta$ is essentially determined by the geometry of the device, while $E_J/E_C$ can be tuned in situ by the external magnetic flux up to a maximum value fixed by the device design.

\begin{figure}
    \centering
        \includegraphics[width=0.7\columnwidth]{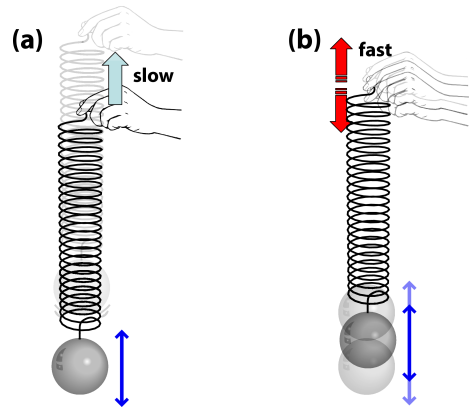}
    \caption{(Color online) Mechanical analogy illustrating the fundamental difference between dc and ac response. (a) For a slow (adiabatic) change in the suspension point of the oscillator, the oscillator mass is displaced but the oscillator frequency does not change.  (b) For an ac drive at resonance, energy is forced into or extracted from the system. Quantum mechanically, this leads to the coupling between the transmon and the cavity field.
    \label{fig:osc}}
\end{figure}

This result is quite remarkable. While the sensitivity of the transmon spectrum to the dc component of $n_g$ \emph{decreases} exponentially, the ac response to the oscillating cavity field \emph{increases} in a power-law fashion. In other words, the charge dispersion and the magnitude of ac coupling are completely disparate.
We can illustrate the fundamental difference between dc and ac response by the following intuitive picture, see Fig.~\ref{fig:osc} \cite{footnotenew}. For large $E_J/E_C$, the transmon can be interpreted as a harmonic oscillator in the charge basis, with its quadratic potential centered at $n=n_g$. Charge noise typically occurs at low frequencies so that it can be treated as an adiabatic displacement of the oscillator potential (dc response). In the general case, this leads to adiabatic changes of the qubit frequency and hence dephasing. However, for a harmonic oscillator the frequency remains unchanged, despite the significant change of the oscillator state under displacement. Thus, the charge dispersion vanishes, and dephasing is exactly eliminated. Remarkably, the transmon approaches this ideal point while retaining sufficient anharmonicity. On the other hand, the question of ac response corresponds to driving the oscillator at its resonance frequency. Classically, the drive transfers energy into the oscillator; quantum-mechanically, it induces transitions between different oscillator states leading to the coupling. 
This illustrates that strong coupling and zero (or exponentially small) charge dispersion are in fact not contradictory.
This is the central point of the transmon:  it is highly polarizable and responds strongly to electric fields at all frequencies.  Just as for a harmonic oscillator however, the adiabatic response to low frequency fields does not lead to changes in the transition frequencies.  Unlike the harmonic oscillator though, the transmon remains moderately anharmonic.

\subsection{Control and readout: The dispersive limit\label{sec:dispersive}}
It has been demonstrated that coherent control and readout of CPB qubits can be achieved by operating the system in the dispersive limit \cite{blais1,wallraff}.  Qubit operations are implemented by means of microwave pulses, readout corresponds to a measurement of the phase or amplitude of the transmitted radiation of a microwave drive field.  In the following, we demonstrate that these concepts may be directly transferred to the transmon: Readout and control of the transmon work exactly the same way as for the CPB.

In the dispersive limit, the detunings $\Delta_i=\omega_{i,i+1}-\omega_r$ between transmon and cavity are large, i.e.~in particular $g_{01}/\abs{\Delta_0}\ll1$, $g_{01}/\abs{\Delta_0+\alpha}\ll1$. In this case, we can eliminate the cavity-qubit interaction to lowest order by a canonical transformation, see Appendix \ref{app1} for details. It is important to note that due to the reduced anharmonicity, virtual transitions through excited transmon states have to be taken into account. Only after this can we restrict the transmon Hilbert space to the ground state and first excited state. This procedure leads to the following effective Hamiltonian
\begin{align}\label{effham}
\hat{H}_\text{eff}
= &\frac{\hbar\omega_{01}'}{2}\hat{\sigma}_z +\left(\hbar\omega_r'+ \hbar\chi \hat{\sigma}_z\right) \hat{a}^\dag \hat{a}. %\\\nonumber
%+ \hbar\lambda(\sigma^-+\sigma^+)a^\dag a
\end{align}
Here, the primes signal parameter renormalizations: both the qubit transition frequency and the resonance frequency of the cavity get renormalized due to the interaction, $\omega_r' = \omega_r -\chi_{12}/2$ and $\omega_{01}'=\omega_{01}+\chi_{01}$. (The definitions of the partial dispersive shifts $\chi_{ij}$ will be discussed below.)

The crucial point of Eq.~\eqref{effham} is that the form of this Hamiltonian is identical to the dynamical Stark-shift Hamiltonian encountered for a CPB coupled to a transmission line resonator\cite{blais1, gambetta}. Remarkably, despite its reduced anharmonicity the transmon behaves in a way quite similar to a CPB when operated in the dispersive regime. This is very convenient as it implies that control and readout techniques previously developed for CPBs can be transferred to the transmon regime. Specifically, the readout proceeds by subjecting the cavity to a microwave field close to its resonance frequency. The ac Stark effect causes a dispersive shift of the resonator frequency depending on the qubit state. Consequently, a measurement of the phase or amplitude of the transmitted field is sufficient to infer the state of the qubit. We stress that this measurement is very different from measurements of the quantum capacitance $\sim\partial E_{i}^2/\partial n_g^2$ \cite{duty}, which would fail due to the exponentially small charge dispersion in the transmon regime.

The only difference between transmon and CPB regards the effective dispersive shift $\chi$ in Eq.~\eqref{effham}. It is given by
\be\label{chi-full}
\chi = \chi_{01} - \chi_{12}/2,
\ee
with
\be\label{chieq}
\chi_{ij}\equiv \frac{g_{ij}^2}{\omega_{ij}-\omega_r}
\ee
and $\omega_{ij}=\omega_j-\omega_i$. We emphasize that these expressions can be obtained either through the canonical transformation presented in  Appendix \ref{app1}, or alternatively, through a straightforward application of second-order perturbation theory. The latter can be understood in terms of pairwise level repulsions between coupled levels as indicated in Fig.~\ref{fig:melements}(b). 

In contrast to the CPB case, the transmon's dispersive shift consists of two contributions which enter with different signs, and which for a pure harmonic oscillator exactly cancel each other. The partial cancellation for the transmon system with low anharmonicity is compensated by the increase in the coupling strength $g$. As a result, the magnitude of the transmon's effective dispersive shift is comparable to that of a CPB.

 As we will show now, the contribution of two terms to the dispersive shift also leads to interesting new physics beyond the usual ac-Stark effect for the two-level case -- in particular, negative dispersive shifts as well as significantly increased positive shifts, depending on the detuning of the qubit.
Using Eq.~\eqref{couplings}, the dispersive frequency shifts between neighboring transmon states can be written as
\be\label{chiij}
\hbar\chi_{i,i+1}=\frac{(2\beta eV^0_\text{rms})^2}{\hbar\Delta_i}\abs{\boket{i}{\hat{n}}{i+1}}^2
\ee
By combining Eqs.~\eqref{n10} and \eqref{chiij}, we obtain the following asymptotic expression for the dispersive frequency shift from Eq.~\eqref{chi-full}, valid for $E_J/E_C\gg1$:
\be\label{chi-as}
\hbar\chi \simeq -(\beta eV^0_\text{rms})^2\left(\frac{E_J}{2E_C}\right)^{1/2}\frac{E_C}{\hbar\Delta_0(\hbar\Delta_0 - E_C)}.
\ee
\begin{figure}
    \centering   
        \includegraphics[width=0.8\columnwidth]{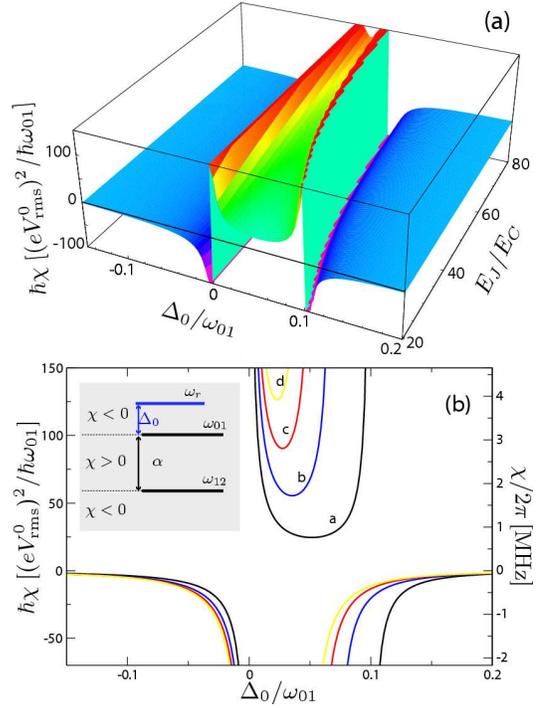}       
    \caption{(Color online) (a) Dispersive frequency shift $\chi$ as a function of detuning $\Delta_0$ between cavity and transmon and energy ratio $E_J/E_C$ (top panel), for fixed transition frequency $\omega_{01}$, as well as $\beta=1/2$ and $n_g=1/2$. (b) Cross-sections of the 3d plot for $E_J/E_C=20\,(a)$, $40\,(b)$, $60\,(c)$, $80\,(d)$. The right-hand side $y$ axis shows the dispersive shift in MHz for the representative parameters chosen in the text. In the general case, the universal axis for $\hbar\chi$ (left-hand side) can be translated into MHz with the conversion factor $\frac{(g_{01}/2\pi)^2}{(\omega_{01}/2\pi)\abs{\boket{0}{\hat{n}}{1}}^2}$.
     Inset: Level configuration and regimes for the dispersive frequency shift.
    \label{fig:chi}}
\end{figure}
The full expression \eqref{chi-full} is plotted in Fig.~\ref{fig:chi}. Intriguingly, the interplay of the contributions from $\chi_{01}$ and $-\chi_{12}/2$ leads to three distinct regions in the $(\Delta_0$, $E_J/E_C)$ plane. These are separated by the poles of Eq.~\eqref{chi-as} [$\hbar\Delta_0=0$, $\hbar\Delta_0=E_C$] where the dispersive limit breaks down. The regions are characterized by differing signs for $\chi$ [see inset of Fig.~\ref{fig:chi}(b)]: (i), (ii) for negative detunings, $\Delta_0<0$, and positive detunings exceeding the absolute anharmonicity, $\Delta_0 > E_C$, the dispersive frequency shift is negative; (iii) for small positive detunings $0<\Delta_0<E_C$, which place the cavity frequency between the transition frequencies $\omega_{01}$ and $\omega_{12}$, $\chi$ becomes positive and obtains comparatively large values. After its special location in parameter space, we name this regime the ``straddling regime.'' The comparatively large dispersive shift renders this region particularly interesting for qubit operations and readout.

It is not completely obvious from the outset that the straddling regime can be occupied without either violating the dispersive constraint $g_{01}/\abs{\Delta_0},g_{01}/\abs{\Delta_0 +\alpha} \ll1$, or losing strong coupling which requires $g\gg \Gamma=\max\{\kappa,\gamma\}$. Here, $\kappa$ and $\gamma$ denote the cavity and qubit decay rates, respectively.  In the following, using realistic values for the relevant parameters, we demonstrate that the straddling regime can indeed be accessed in the strong-coupling and dispersive regime, and that it can lead to larger overall dispersive shifts $\chi$.

Based on data from a recent experiment \cite{schuster},  decay rates of the order of $\Gamma/2\pi=2\,\text{MHz}$ are achievable. This is much lower than the strong coupling $g/2\pi\simeq 100\,\text{MHz}$ demonstrated in experiments with the qubit operated at a transition frequency of $\omega_{01}/2\pi\simeq7\,\text{GHz}$. We argue that the charging energy cited in Ref.~\cite{schuster} can be be further lowered to values of the order of $E_C/h=300\,\text{MHz}$. 

First, we consider the case of negative detuning, $\Delta_0/2\pi=-150\,\text{MHz}$, i.e.\ the cavity frequency is higher than all transmon transition frequencies. Assuming a coupling strength of $g_{01}/2\pi=20\,\text{MHz}$, the conditions for strong coupling and the dispersive limit are met, and we obtain a negative dispersive frequency shift of $\chi\approx-1.4\,\text{MHz}$. This should be contrasted with the situation of positive detuning $\Delta_0/2\pi=150\,\text{MHz}$ in the straddling regime, i.e.\ the cavity frequency is located between the transition frequencies $\omega_{01}$ and $\omega_{12}$. Very remarkably, this sign change of the detuning leads to an \emph{increase} of the dispersive frequency shift to $\chi=3.4\,\text{MHz}$. We have also confirmed the validity of the dispersive straddling regime by a numerical diagonalization of the full Jaynes-Cummings Hamiltonian.

%and a coupling of $g/2\pi\simeq 100\,\text{MHz}$. Clearly, the conditions for strong coupling and the dispersive limit are met. Identifying $g=g_{01}$, we obtain a negative dispersive frequency shift of $\chi\approx-0.9\,\text{MHz}$. This should be contrasted with the situation of small detuning $\Delta_0/2\pi=150\,\text{MHz}$. Choosing a smaller coupling of $g/2\pi=20\,\text{MHz}$, we ensure that the dispersive limit still applies. Very remarkably, this reduction of the coupling and the smaller detuning lead to an \emph{increase} of the dispersive frequency shift to $\chi=5.3\,\text{MHz}$. We have also confirmed the validity of the dispersive straddling regime by a numerical diagonalization of the full Jaynes-Cummings Hamiltonian.

\section{Estimates for the transmon's relaxation time ($T_1$)\label{sec:relaxation}}
We have argued that a main advantage of the transmon as compared to other existing solid-state qubits is its remarkable insensitivity to charge noise. In this and the following section, we investigate in detail the sensitivity of the transmon to various noise channels and discuss its performance in terms of the projected relaxation and dephasing times.

\subsection{Relaxation by spontaneous emission}
The fact that the transmon qubit couples to the electromagnetic field inside the transmission line resonator indicates that radiative decay of the transmon is one inevitable relaxation channel. We can estimate the order of magnitude of the resulting $T_1$ by a simple semi-classical argument, which agrees with the quantum result from Fermi's golden rule. The average power emitted into free space from an electric dipole with dipole moment $d$, oscillating at angular frequency $\omega$ is given by
\be
P=\frac{1}{4\pi\epsilon_0}\frac{d^2\omega^4}{3c^3},
\ee
see e.g.~\cite{jackson}.
We can obtain an estimate for the transmon's dipole moment from the distance a Cooper pair travels when tunneling between the two superconducting islands. In a recent experiment approaching the transmon limit \cite{schuster}, this distance is of the order of $L\sim15\,\mu\text{m}$. Hence, as an estimate for the dipole moment we obtain $d=2eL$. 
As a result, the decay time for the excited transmon level due to emission of radiation is given by
\be
T_1^\text{rad} = \hbar\omega_{01}/P = \frac{12\pi\epsilon_0\hbar c^3}{d^2 \omega_{01}^3}.
\ee
For a realistic qubit frequency $\omega_{01}/2\pi=8\,\text{GHz}$ this leads to relaxation times of the order of $0.3\,\text{ms}$ and a corresponding $Q$ factor of $10^{7}$.

\subsection{The Purcell Effect} 
When a system is placed inside a resonator, its spontaneous emission rate is altered.  This effect is known as the Purcell effect \cite{purcell}. It has been observed in microwave cavities using Rydberg atoms \cite{goy}, and recently also in electrical circuits \cite{houck}.  For the transmon coupled to a transmission line resonator, the same effect will occur and each transmon level will experience a different change to its spontaneous relaxation rate.  In this section we compute these rates.

The simplest way to obtain estimates for the decay rates is to apply Fermi's golden rule to the Hamiltonian that describes the interaction of the resonator with its bath. This Hamiltonian is  
\be
H_{\rm \kappa}= \hbar\sum_k \lambda_k \left[\hat{b}^{\dg}_k \hat{a}  + \hat{a}^{\dg} \hat{b}_k\right] \ee
 where $\hat{b}_k$ and $\hat{b}_k^\dg$ are the bath operators for mode $k$, and $ \lambda_k$ determines the coupling strength of the resonator to this bath mode. 
 Using Fermi's golden rule and performing a continuum limit, the rate for a transition from an eigenstate $\ket{i}$  of the Hamiltonian \eqref{hamiltonian-before2} to the final state $\ket{f}$ is  
\be
\gamma_\kappa^{(f,i)} = \frac{2\pi}{\hbar} p(\omega_k) \abs{\boket{1_{k},f}{ \hbar\sum_{k'}\lambda_{k'}\left[\hat{b}^{\dag}_{k'} \hat{a} +
\hat{a}^\dg \hat{b}_{k'}\right]}{0,i}}^2,
\ee
here exemplified for the case of the loss of one photon with energy $\hbar\omega_k=E_i-E_f$ to the bath. The reservoir's density of states at energy $\hbar\omega_k$ is denoted by $p(\omega_k)$.  Defining $\kappa = 2\pi\hbar p(\omega_k) |\lambda_k|^2$, this can be written as  
\be
\gamma_\kappa^{(f,i)} =  \kappa \left| \bra{f} \hat{a} \ket{i} \right|^2.
\ee
In the resonant limit of zero detuning, the system's lowest excitation consists of an equally weighted superposition of the single photon state and the excited qubit state.  Consequently, we obtain the standard result that  the resonator's bath will relax this excited state at rate $\kappa/2$.

By contrast, in the dispersive limit the transmon states acquire only a small photonic component (and photons only a small qubit component), which can be obtained perturbatively from Eq.~\eqref{hamiltonian-before2},
\begin{align}
\overline{\ket{n,j}}=\ket{n,j} &+ \frac{\sqrt{n}g_{j,j+1}}{\omega_{j,j+1}-\omega_r}\ket{n-1,j+1}\\\nonumber &-\frac{\sqrt{n+1}g_{j,j+1}}{\omega_{j-1,j}-\omega_r}\ket{n+1,j-1}.
\end{align}
As a result, the spontaneous emission rate for the excited qubit state $\overline{\ket{0,1}}$ will acquire the additional Purcell contribution
\be
\gamma_\kappa^{(0,1)}= \kappa\frac{g^2_{01}}{\Delta_0^2}. 
\ee
Similarly, in the absence of photons in the cavity the higher transmon levels will see a Purcell-induced relaxation rate of 
\be
\gamma_\kappa^{(i,i+1)} =  \kappa \frac{g^2_{i,i+1}}{(\omega_{i,i+1}-\omega_r)^2}.
\ee
For a cavity with lifetime $1/\kappa=160\,\text{ns}$ ($Q=10^4$) and $g_{01}/\Delta_0=0.1$,  the Purcell effect leads to a $T_1$ contribution of $16\,\mu\text{s}$.

\subsection{Dielectric losses}
Dielectric losses from insulating materials, especially amorphous SiO$_2$, have recently been marked as a potentially crucial channel of relaxation in superconducting qubits \cite{martinis2}. While substrates such as crystalline Si and sapphire offer favorably low loss tangents $\tan \delta = \rre \epsilon_r /\iim \epsilon_r$ of the order of $10^{-6}$ and $10^{-8}$ at cryogenic temperatures, amorphous SiO$_2$ has been found to exhibit loss tangents as large as $5\times10^{-3}$ \cite{martinis2}. Such dielectric losses affect the electric fields associated with the qubits and cause energy relaxation with a rate proportional to $\tan \delta$.

These findings clearly call for a cautious choice of materials when designing superconducting qubits. In addition, we point out that the participation ratio of the different materials present in the immediate vicinity of the actual qubit will play a similarly crucial role. In particular, in the transmon design the shunting capacitance $C_B$ offers the possibility to accumulate a large percentage of the electric fields in a well-controlled spatial region with favorable substrates, as opposed to storing them in the less well-defined Josephson junction region. As a result, we expect good performance of the transmon in terms of robustness with respect to dielectric losses. An additional significant advantage of the transmon geometry is that it can be fabricated with single layer processing with no deposited dielectric layers that might cause large losses.

\subsection{Relaxation due to quasiparticle tunneling\label{sec:qprelax}}
\begin{figure}
    \centering
        \includegraphics[width=0.8\columnwidth]{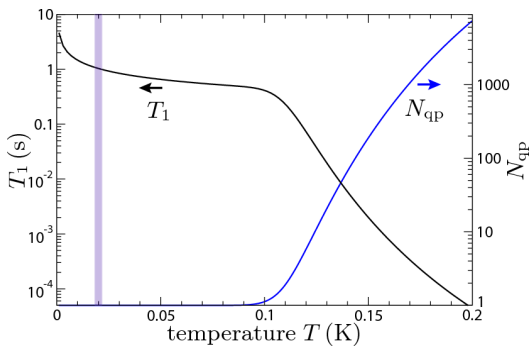}
    \caption{Number of quasiparticles and contributions to the relaxation time $T_1$ due to inelastic quasiparticle tunneling as a function of temperature at $E_J/E_C=60$. From this estimate, tunneling of quasiparticles is not expected to limit the performance of the transmon at cryogenic temperatures. In typical dilution refrigerator experiments, (phonon) temperatures are of the order of $20\,$mK, marked in the plot by a vertical bar.
\label{fig:quasiparticles}}
\end{figure}

The presence of quasiparticles in the system, due to an overall odd number of electrons or thermal breaking of Cooper pairs, leads to both relaxation and dephasing in qubits based on Josephson junctions \cite{lutchyn1,lutchyn2}. Following the arguments by Lutchyn at el.\ \cite{lutchyn2} we may estimate the resulting decoherence rates for the transmon system. With the total number of conduction electrons given by $N_e=n V$, $n$ and $V$ denoting the conduction electron density and the metal volume respectively, the number of quasiparticles may be obtained as
\be
N_\text{qp}=1+\frac{3\sqrt{2\pi}}{2}N_e \frac{\sqrt{\Delta\,k_BT}}{E_F} e^{-\Delta/k_BT},
\ee
valid for temperatures small compared to the superconducting gap $\Delta$. We have assumed there is thermal breaking of Cooper pairs, as well as one unpaired electron, which could naturally arise if the finite volume of the qubit contains an odd number of electrons. The rate of tunneling for one quasiparticle across the junction is given by $\Gamma_\text{qp}=\delta g_T/4\pi\hbar$ \cite{lutchyn2}, where $\delta=1/\nu V$ is the mean level spacing of the reservoir, $\nu$, $V$ its density of states and volume and $g_T$ the junction conductance measured in units of $e^2/h$. Generalizing the expressions from \cite{lutchyn2} to the transmon regime, we obtain for the full relaxation rate due to quasiparticle tunneling:
\begin{align}\label{qprate}
\Gamma_1&=1/T_1 \simeq \Gamma_\text{qp}N_\text{qp} \sqrt{\frac{k_B T}{\hbar\omega_{01}}}\abs{\bket{g,n_g\pm 1/2}{e,n_g}}^2
%\nonumber\\
%&=\frac{g_T}{4\pi\hbar\nu}\left[ \frac{1}{V} + n \frac{\Delta}{E_F}e^{-\Delta/k_BT}\right] \sqrt{\frac{k_B %T}{\hbar\omega_{01}}}\abs{\bket{g,n_g\pm 1/2}{e,n_g}}^2
\end{align}
The matrix element is the Franck-Condon factor which accounts for the shake-up of the transmon collective mode due to tunneling of one quasiparticle in the sudden approximation, and in the $E_J/E_C\gg1$ limit becomes independent of $n_g$. For our rough estimate, we disregard the issue of possible nonequilibrium quasiparticle distributions and use typical parameter values ($V=300\times5\times0.1\,\mu\text{m}^3$, $g_T=1$, $T=20\,\text{mK}$, aluminum: $E_F=11.7\,\text{eV}$, $n=18.1\times10^{22}\,\text{cm}^{-3}$, $\nu=3n/2E_F$) and obtain a relaxation time due to quasiparticle tunneling of the order of $T_1\sim1\,\text{s}$. It is important to note that the increased volume of the transmon does \emph{not} lead to smaller $T_1$ in the quasiparticle channel. In fact, the matrix element in Eq.~\eqref{qprate} leads to an increase of $T_1$ for large $E_J/E_C$ because the quantum fluctuations of the coherent charge grow larger relative to the charge displacement caused by the quasiparticle tunneling. The projected temperature dependence of the $T_1$ contribution due to quasiparticle tunneling is depicted in Fig.~\ref{fig:quasiparticles}.
The conclusion we draw from Fig.~\ref{fig:quasiparticles} is that below 100\,mK quasiparticles should not lead to significant contributions to relaxation in the transmon. This result should be rather robust to the actual number of quasiparticles present in the limit of $T\to0$, since a relevant decrease in $T_1$ is expected only when the number of quasiparticles reaches several thousands.

\subsection{Relaxation due to flux coupling}
\begin{figure}[t]
    \centering
        \includegraphics[width=0.55\columnwidth]{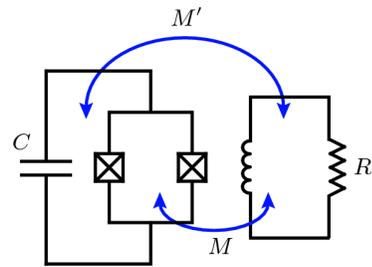}
    \caption{(Color online) Model for the estimate of relaxation times due to flux coupling, describing (i) flux coupling between the transmon's SQUID loop and the external flux bias with mutual inductance $M$, and (ii) flux coupling between the transmon circuit and an external flux bias circuit via the mutual inductance $M'$.
    \label{fig:fluxcoupling}}
\end{figure}

The coupling of the transmon to an external magnetic flux bias allows for an in situ tuning of the Josephson coupling energy, but also opens up additional channels for energy relaxation: (i) there is an intentional coupling between the SQUID loop and the flux bias (allowing for the $E_J$ tuning) through a mutual inductance $M$; (ii) in addition, the entire transmon circuit couples to the flux bias via a mutual inductance $M'$; see Fig.~\ref{fig:fluxcoupling}. Here, we provide simple order of magnitude estimates of the corresponding relaxation times.
For the estimate of relaxation rates due to the mechanism (i), we assume that the overall flux applied to the SQUID ring can be decomposed into the external flux and a small noise term, i.e.\ $\Phi=\Phi_e+\Phi_n$ with $\Phi_n\ll\Phi_e$. Then, a Taylor expansion of the Josephson Hamiltonian \eqref{josham} yields 
\be
\hat{H}_J \to \hat{H}_J + \Phi_n \hat{A}
\ee
where
\begin{align}
\hat{A} &= \frac{\partial \hat{H}_J}{\partial \Phi}\bigg|_{\Phi_e} \\\nonumber
&= E_{J\Sigma}\frac{\pi}{\Phi_0}\left[ \sin \left( \frac{\pi\Phi_e}{\Phi_0} \right) \cos \hat{\varphi} -  d\cos \left( \frac{\pi\Phi_e}{\Phi_0} \right) \sin \hat{\varphi}\right].
\end{align}
As in Section \ref{sec:flux}, $E_{J\Sigma}=E_{J1}+E_{J2}$ denotes the total Josephson energy and $d=(E_{J1}-E_{J2})/E_{J\Sigma}$ parameterizes the junction asymmetry.
Treating the noise perturbatively, one can relate the relaxation rate to the noise power spectrum, see e.g.\ \cite{schoelkopf},
\begin{align}
\Gamma_1&=\frac{1}{T_1}=
%\frac{|\boket{1}{\hat{A}}{0}|^2}{\hbar^2}S_{\Phi_n}(\omega_{01})
\frac{1}{\hbar^2}|\boket{1}{\hat{A}}{0}|^2{M}^2 S_{I_n}(\omega_{01}).
%\\\nonumber
%&=\frac{E_{J\Sigma}^2\pi^2M^2}{\Phi_0^2\hbar^2}S_{I_n}(\omega_{01})\abs{\sin \left( \frac{\pi\Phi_e}{\Phi_0} \right)
%\boket{1}{\cos\hat{\varphi}}{0}- d \cos \left( \frac{\pi\Phi_e}{\Phi_0} \right)
%\boket{1}{\sin\hat{\varphi}}{0}}^2.
\end{align}
Here, we have made use of the connection between flux noise and current noise determined by the mutual inductance, $S_{\Phi_n}(\omega)={M^2}S_{I_n}(\omega)$. At low temperatures $k_BT \ll \hbar\omega_{01}$ the current quantum noise is given by $S_{I_n}(\omega)=2\Theta(\omega)\hbar\omega/R$. For a typical junction asymmetry of $10\%$ and realistic device parameters ($E_J=20\,\text{GHz}$, $E_C=0.35\,\text{GHz}$, $M=140\Phi_0/A$, $R=50\,\Omega$) we obtain relaxation times ranging between $20\,\text{ms}$ and $1\,\text{s}$, where the maximum (minimum) $T_1$ is reached for an integer (half-integer) number of flux quanta threaded through the SQUID loop.

For the decay channel (ii), we may model the entire transmon circuit by a simple $LC$ oscillator with $L\approx \hbar^2/4e^2E_J$ and $C\approx e^2/2E_C$. Classically, the charge then oscillates according to $Q(t)=Q_0 \cos \omega t$ with oscillator frequency $\omega=1/\sqrt{LC}$. Assuming that the energy stored in the oscillator is of the order of one energy quantum $\hbar\omega$, we obtain $Q_0=\sqrt{2C\hbar\omega}$ and $I(t)=-I_0\sin \omega t$ with $I_0=\omega\sqrt{2C\hbar\omega}$. Through the mutual inductance, this oscillating current induces a voltage $V_\text{ind}(t) = V_0 \sin \omega t$ in the flux bias circuit, where $V_0=M'\omega^2 \sqrt{2C\hbar\omega}$. The environmental $R\sim50\Omega$ impedance will dissipate the average power $P=V_0^2/2R$, which allows for the following estimate of the relaxation time:
\be
T_1\simeq \frac{\hbar\omega}{P} = \frac{R}{{M'}^2\omega^4C}=\frac{RC}{\eta^2}
\ee
where $\eta=M'/L$ measures the effective coupling strength in units of the Josephson inductance. It is crucial to note that for the particular case of a SQUID loop and a flux bias line exactly centered in the middle of the transmission line resonator [cf.\ Fig.\ \ref{fig:circuit-diagram}(b)], the mutual inductance $M'$ identically vanishes for symmetry reasons and relaxation via this channel would not occur. However, when realizing the flux bias line with a coplanar waveguide, it is natural to displace the line in order to maximize coupling to the SQUID loop. The resulting mutual inductance can be estimated and we obtain values of the order of $M'=10\Phi_0/\text{A}$. Using realistic numbers for the Josephson coupling, charging energy, and the environmental impedance [$E_J=20\,\text{GHz}$, $E_C=0.35\,\text{GHz}$, $R=50\Omega$] we obtain relaxation times of the order of $70\,\text{ms}$. Therefore, relaxation due to flux coupling is unlikely to limit the performance of the transmon system. This can be explained by the smallness of the effective coupling $\eta$, which for the parameters above is of the order of $10^{-6}$.

\subsection{Coupling to spurious modes, balancing, and other noise sources}
It is a common phenomenon that measured relaxation (and dephasing) times tend to be shorter than theoretically predicted, indicating that the microscopic origin of decoherence is not yet well understood in all cases and that additional, unknown noise channels participate. Various candidates for this can be discussed, such as coupling to spurious resonator modes and pinning and unpinning of vortices etc. In addition, phonon emission due to bulk and interface piezoelectricity in solid-state qubits has recently been pointed out as another possible source for energy relaxation \cite{ioffe}. While bulk piezoelectricity can be avoided by selecting materials with inversion symmetry, interface piezoelectricity will always be present. With a Rayleigh velocity of $v_R=4.92\,\text{nm}/\text{ps}$ for surface acoustic waves (SAWs) in Si \cite{hurley}, one obtains acoustic wavelengths of the order of $\lambda_R\sim 1\,\mu\text{m}$ at typical qubit frequencies. Consequently, there is a significant mismatch between the acoustic wavelength and the overall lateral size of the transmon, $\lambda_R\ll  30\,\mu\text{m}$. This makes an efficient coupling to surface acoustic wave modes less likely. In addition, well-defined mesas with step heights exceeding the evanescent wave depth will act as mirrors for SAWs, which may be further exploited in the device design to minimize coupling to SAW modes. An accurate estimate of this coupling will require a detailed modeling of the transmon's capacitor shapes.
 Thus, obtaining quantitative estimates for the decoherence rates in these additional channels may not always be simple. 
 Ultimately, the experiment will determine whether they play any role in decohering the transmon qubit.

We finally emphasize that the design shown schematically in Fig.~\ref{fig:circuit-diagram}(b) may require special care due to the possibility of imbalance in the coupling to the transmission line resonator. Due to its location between bottom line and center pin, the transmon will tend to couple to a spurious slotline mode of the resonator in which bottom and top line are out of phase. It is important to note that this issue is eliminated by a design with top/bottom symmetry. Alternatively, a tailoring of the relevant capacitances in the spirit of a Wheatstone bridge can compensate for the problem, thus effectively balancing the asymmetry between top and bottom line that is induced by the geometric layout.

\section{Estimates for the transmon's dephasing time ($T_2$)\label{sec:dephasing}}
Generally, the origin of qubit dephasing can be understood in terms of fluctuations of the qubit transition frequency due to its coupling  to the environment. This can be described as noise in the external parameters $\lambda_i$ of the Hamiltonian. In the present case of the transmon, prominent noise sources are the charge fluctuations as well as the fluctuations in the critical current and the magnetic flux. We first review the general formalism appropriate to the treatment of small fluctuations around the controlled dc value of the external parameters and turn to the investigation of the individual noise channels in the subsequent subsections.

Formally, the qubit Hamiltonian may be written as
\be
\hat{H}_\text{q}= \frac{1}{2}\sum_{u=x,y,z} h_u(\{\lambda_i\})\hat{\sigma}_u,
\ee
and each external parameter can be decomposed into its (controlled) dc value and fluctuations around it, $\lambda_i=\lambda_{i}^{(0)}+\delta\lambda_i$.  Following Refs.\ \cite{makhlin2} and \cite{ithier}, the case 
of weak fluctuations can be addressed by a  Taylor expansion with respect to $\delta\lambda_i$. To lowest order we have
\be\label{hexp}
\hat{H}_\text{q}= \frac{\hbar \omega_{01}}{2}\hat{\sigma}_z + \frac{1}{2}\sum_{j}\sum_{u=x,y,z}  \frac{\partial h_u(\{\lambda_i\})}{\partial \lambda_j}\delta\lambda_j \hat{\sigma}_u +\mathcal{O}(\delta\lambda^2),
\ee
where all derivatives are evaluated at the dc values $\lambda_i^{(0)}$. The fluctuations $\delta\lambda_i$ generally result in two distinct effects: (i) For sufficiently low frequencies, the fluctuations in the $\sigma_z$ component can be treated within the adiabatic approximation. In this case, they cause random shifts of the transition frequency of the qubit, leading to pure dephasing (time scale $T_2$). (ii) Higher frequencies will break the adiabatic approximation and induce transitions between qubit states (energy relaxation, time scale $T_1$). 
%In addition, fluctuations acting in the $\sigma_x$ and $\sigma_y$ space will always lead to transitions.
  Here, we focus on the dephasing aspect (i), i.e.\ we consider low-frequency noise in the $\sigma_z$ component. 

It is convenient to define $T_2$ via the law for the decay of the off-diagonal density matrix elements, which is given by
\be
\rho_{01}(t)=e^{i\omega_{01}t}\left\langle e^{ -i \int_0^t dt' v(t') } \right\rangle,
\ee
where $v(t)=\sum_j \frac{\partial h_z(\{\lambda_i\})}{\hbar \partial \lambda_j}\delta\lambda_j$. Assuming Gaussian noise, it is straightforward to carry out the noise average, and the result can be expressed in terms of the noise power 
\be
S_v(\omega)=\int_{-\infty}^\infty d\tau \langle v(0)v(\tau) \rangle e^{-i\omega \tau}=\sum_j \frac{\partial h_z(\{\lambda_i\})}{\hbar \,\partial \lambda_j}S_{\lambda_j}(\omega),
\ee
which yields \cite{makhlin2,ithier},
\begin{align}%\nonumber
\rho_{01}(t)%&=e^{i\omega_{01}t}\exp\left[ -\frac{1}{2} \int_0^t dt_1 \int_0^t dt_2 \left\langle v(t_1)v(t_2) \right\rangle\right]\\
&=e^{i\omega_{01}t}\exp\left[ -\frac{1}{2}\int_{-\infty}^\infty \frac{d\omega}{2\pi} S_v(\omega)\frac{\sin^2(\omega t/2)}{(\omega/2)^2}\right].\label{r01}
\end{align}
As a consequence, the resulting dephasing critically depends on the magnitude of the autocorrelation time $t_c$ of the noise. For correlation times small compared to the typical acquisition time, $t_c\ll t$, the dephasing follows an exponential law
\be
\rho_{01}(t)\simeq e^{i\omega_{01}t}\exp\left[-\frac{1}{2}\abs{t}S_v(\omega=0)\right].
\ee
The corresponding lineshape is Lorentzian (\emph{homogeneous broadening}) and the dephasing time $T_2\simeq 2/S_v(\omega=0)$. Note that this expression is valid for noise spectra with a regular low-frequency behavior. For spectra singular at $\omega=0$, the full equation \eqref{r01} has to be invoked. In particular, this applies to the case of $1/f$ noise,
\be
S_{\lambda_i}(\omega)=\frac{2\pi A^2}{\abs{\omega}^\mu},
\ee
 which has been identified as the typical noise spectrum for various noise channels. In experiments, the scaling exponent $\mu$ in $\omega^{-\mu}$ is typically in a range $0.8<\mu<1.3$. For simplicity, we assume $\mu=1$ in the following. The parameter $A$ determines the overall amplitude of the fluctuations and will be specified for the various noise sources below; see also Table \ref{tab1}. Denoting the 
 infrared and ultraviolet cutoffs by $\omega_i$ and $\omega_u$,  this leads to 
\be
\rho_{01}(t)\simeq e^{i\omega_{01}t}\exp\left[-\frac{A^2}{\hbar^2}\left(\frac{\partial h}{\partial \lambda_i}\right)^2 t^2 \abs{\ln \omega_i t}\right],\label{decaylaw}
\ee
valid for $\omega_i t \ll 1$ and $\omega_u t\gg1$, see e.g.\ Refs.\ \cite{ithier,cottet3,makhlin2}. 
%In this case, $T_2$ depends on the infrared cutoff in a non-algebraic way.

\subsection{Charge noise}
We now relate the differential charge dispersion $\partial E_{01}/\partial n_g$  to the sensitivity of the qubit with respect to charge noise. 
The charge noise observed for transmon devices in experiments \cite{houck2} indicates that large fluctuations of the offset charge only occur on time scales exceeding the typical acquisition time of a single experiment. Additionally, small fluctuations may persist within each single shot. Both aspects contribute to dephasing and are investigated in the following.

Small fluctuations can be treated in terms of Eq.~\eqref{decaylaw} with the external parameter given by $\lambda_i\to n_g$.  A rough estimate of the resulting dephasing time $T_2$ is obtained by substituting the logarithmic contribution by a constant, such that 
\be\label{charget2}
T_2\sim \frac{\hbar}{A}\abs{ \frac{\partial E_{01}}{\partial n_g}}^{-1}\simeq\frac{\hbar}{A\pi\abs{\epsilon_1}}.
\ee
Using Eq.~\eqref{bgcurvature}, we indeed find an exponential increase of $T_2$ for large $E_J/E_C$. Hence, the qubit becomes essentially immune to charge noise in this limit. Using realistic parameter values [$E_J=30\,\text{GHz}$, $E_C=0.35\,\text{GHz}$, $A=10^{-4}$ \cite{zorin}] we obtain a dephasing time of the order of $T_2\sim8\,\text{s}$.

We now turn to the investigation of slow charge fluctuations with large amplitudes, which cannot be treated within a perturbative scheme. In this case, we explicitly write the qubit Hamiltonian as
\be
\hat{H}_\text{q}\approx \frac{1}{2}\left[\hbar\omega_{01}+\frac{\epsilon_1}{2}\cos(2\pi n_g+ 2\pi \delta n_g(t))\right]\hat{\sigma}_z.
\ee
The corresponding decay law of the off-diagonal density matrix element then reads
\begin{align}
&\rho_{01}(t)\\\nonumber
&\simeq e^{i\omega_{01}t}\bigg\langle \exp\left[-i \frac{\epsilon_1}{2\hbar} \int_0^t dt'\,\cos[2\pi( n_g+ \delta n_g(t'))]\right]\bigg\rangle.
\end{align}
For variations slow compared to the typical measurement time, the effective offset charge will vary for different runs but remain constant within each single run, and we can substitute $n_g+\delta n_g(t)$ by a single constant. As the worst-case scenario, we will assume that the effective offset charge randomly switches according to a uniform probability distribution on $[0,1]$. This results in
\begin{align}
\rho_{01}(t)&\simeq e^{i\omega_{01}t}\int_0^1 dn_g\, \exp\left[-i \epsilon_1 t \cos(2\pi n_g)/2\hbar\right]\nonumber\\ &=e^{i\omega_{01}t}\bj_0(\abs{\epsilon_1}t/2\hbar)  .
\end{align}
The envelope of the Bessel function $\bj_0(z)$ asymptotically falls off as $\sqrt{2/\pi z}$. Thus, using the $1/e$ threshold as a measure for the dephasing time, we obtain $T_2\simeq\frac{4\hbar}{e^2\pi\abs{\epsilon_1}}$, which again increases exponentially with $E_J/E_C$. For the parameter values used above, we find a $T_2$ of the order of $0.4\,\text{ms}$. Hence, we find that the increase in the ratio $E_J/E_C$ featured by the transmon leads to an exponential immunization of the device against charge noise. 

For a full appreciation of these numbers, it is useful to contrast our results with the charge dephasing in a regular CPB operated at the charge sweet spot. In this case, noise is eliminated to linear order and second-order contributions dominate. Taking these into account, Eq.~\eqref{charget2} can be generalized and the CPB dephasing time due to charge noise may be expressed as \cite{ithier}
\be\label{2ndorder}
T_2\simeq \abs{ \frac{\pi^2A^2}{\hbar}\frac{\partial^2 E_{01}}{\partial n_g^2}}^{-1}_{n_g=1/2}=\frac{\hbar}{A^2\pi^2}\frac{E_J}{64E_C^2},
\ee
which for parameter values realized in experiments on CPBs \cite{vion2} leads to dephasing times of the order of $T_2\sim1\,\mu\text{s}$.

\subsection{Flux noise}
Noise in the externally applied flux translates into fluctuations of the effective Josephson coupling energy $E_J$. For simplicity, we consider the symmetric junction case $d=0$. Since in the experiment flux is used to tune the qubit transition frequency via $E_J$ \cite{houck}, we consider both flux noise at and away from the flux sweet spot.  Assuming that the flux noise is sufficiently small, we can employ Eq.~\eqref{decaylaw} and obtain
\begin{align}\label{fluxdeph}
T_2&\simeq \frac{\hbar}{A}\abs{\frac{\partial E_{01}}{\partial \Phi}}^{-1} \\\nonumber
&= \frac{\hbar}{A}\frac{\Phi_0}{\pi}\left(2E_CE_{J\Sigma} \abs{\sin \frac{\pi\Phi}{\Phi_0} \tan \frac{\pi\Phi}{\Phi_0}}\right)^{-1/2},
\end{align}
valid for $E_J\gg E_C$. In this transmon regime, the device is necessarily operated away from points of half-integer numbers of flux quanta where $E_J=0$ and $T_2$ vanishes. For an order of magnitude estimate, we may use a flux bias of $\Phi=\Phi_0/4$ and representative values $A=10^{-5}\Phi_0$ \cite{wellstood},  $E_{J\Sigma}=30\,\text{GHz}$, and $E_C=0.35\,\text{GHz}$. This yields a dephasing time of the order of $T_2\sim 1\,\mu\text{s}$. 

It is important to note that Eq.~\eqref{fluxdeph} results in infinite $T_2$ for an integer number of flux quanta. This is the flux sweet spot \cite{vion2}, where second-order contributions [neglected in Eq.~\eqref{hexp}] dominate. Analogous to Eq.~\eqref{2ndorder}, the dephasing time is given by
\be
T_2\simeq \abs{ \frac{\pi^2A^2}{\hbar}\frac{\partial^2 E_{01}}{\partial \Phi^2}}^{-1}_{\Phi=0}=\frac{\hbar\Phi_0^2}{A^2\pi^4\sqrt{2E_{J\Sigma}E_C} }.
\ee
Using the same representative parameters as above, we obtain for the sweet spot a dephasing time of the order of $T_2\sim 3.6\,\text{ms}$. By comparison, flux dephasing at the double sweet spot in the CPB regime yields a dephasing time of
\be
T_2\simeq \frac{\hbar\Phi_0^2}{A^2\pi^4 E_{J\Sigma}},
\ee
which for realistic parameters gives values of the order of $1\,\text{ms}$.

\subsection{Critical current noise}
A second source of fluctuations of the Josephson energy consists of noise in the critical current, which is believed to be generated by trapping and detrapping of charges associated with spatial reconfigurations of ions inside the tunneling junction \cite{harlingen}. Such rearrangements in the junction directly influence the critical current and hence the Josephson coupling energy $E_J=I_c \hbar/2e$. The corresponding dephasing time for the transmon is obtained as
\begin{align}\label{Icdeph}
T_2&\simeq \frac{\hbar}{A}\abs{\frac{\partial E_{01}}{\partial I_c}}^{-1}
%=\frac{e}{2A}\sqrt{8E_J/E_C} 
= \frac{2\hbar}{\bar{A}E_{01}},
\end{align}
where $\bar{A}=A/I_c$ denotes the dimensionless fluctuation amplitude (independent of $E_J$).
For $A=10^{-6}I_c$ (based on \cite{harlingen}),  $E_{J\Sigma}=30\,\text{GHz}$, and $E_C=0.35\,\text{GHz}$, this results in a dephasing time of the order of $T_2\sim35\,\mu\text{s}$. For comparison, evaluation of the critical current noise in the CPB regime yields $T_2\simeq \frac{\hbar}{\bar{A}E_{01}}$, i.e.\ half the dephasing time expected for the transmon. We remark that based on our estimates, critical current noise is likely to be the limiting dephasing mechanism. 

\subsection{Dephasing due to quasiparticle tunneling}
In principle, the tunneling of quasiparticles not only results in relaxation but also in dephasing. For the CPB regime, $E_J\alt E_C$, Lutchyn et al.\ demonstrated that the dephasing rate may be approximated by $\Gamma_2^\text{qp}\simeq \Gamma_\text{qp}N_\text{qp}$ \cite{lutchyn2}, where we are using the same definitions as in Section \ref{sec:qprelax}. A crucial ingredient in this estimate is the fact that in the charge limit, the transition frequency of the qubit is drastically altered when adding or removing a single charge from the island. As a result, in the CPB regime complete dephasing is achieved by the tunneling of a single quasiparticle. We emphasize that this does \emph{not} hold in the transmon regime, where the charge dispersion is exponentially flat so that transition frequency variations due to a single charge are minimal. Instead, we expect that dephasing due to quasiparticles will mainly be induced by tunneling-induced relaxation processes, described in Section \ref{sec:qprelax}.

\subsection{$E_C$ noise}
Remarkably, as compared to the Cooper pair box the transmon should feature extended relaxation and dephasing times in all noise channels discussed so far. A possible channel for which this does not hold is noise in the charging energy, i.e.\ fluctuations in the effective capacitances of the circuit. So far, there does not seem to be any concrete experimental evidence for this type of noise. In distinction to the critical current, spatial reconfigurations of atoms or groups of atoms inside junctions should only weakly affect the actual capacitances (no exponential dependence). Due to the presence of a large shunting capacitance in the transmon, $E_C$ noise (if existent) could be more important in this system than for the CPB. However, presently there is no evidence for this to be a limiting factor.

\begin{table*}[ht]
    \centering
      \begin{ruledtabular}\squeezetable 
        \begin{tabular}{llll}
        Noise source  && transmon  & CPB \\
        && $E_J/E_C=85$ & $E_J/E_C=1$ \\\hline\hline
        \textbf{dephasing} &$1/f$ amplitude &$T_2$ [ns] &$T_2$ [ns]  \\\hline
        charge & $A=10^{-4}-10^{-3}e$ \cite{zorin}  & 400,000  & \textbf{1,000}$^*$ \\
        flux & $A=10^{-6}-10^{-5}\Phi_0$ \cite{wellstood,yoshihara}  & 3,600,000$^*$ & 1,000,000$^*$ \\
        crit. current & $A=10^{-7}-10^{-6}I_0$ \cite{harlingen} & \textbf{35,000}  & 17,000
        \end{tabular}
        \end{ruledtabular}
        \caption{Comparison of dephasing times for the transmon and Cooper pair box qubits. Contributions to  $T_2$   are theoretical predictions based on Section \ref{sec:dephasing}. Entries in bold face mark the dominant noise channel. For the CPB second-order charge noise at the sweet spot is most likely limiting the performance of the qubit. In contrast, for the transmon dephasing is suppressed to an extent that coherence times should be limited by relaxation ($T_1$) processes only.  All times are given in ns, which is close to the clock cycle used in experiments. Typical qubit frequencies are of the order of $1$--$10\,$GHz, pulse durations usually range between $1$ and $10\,$ns.\\\noindent
        $^*$These values are are evaluated at a sweet spot (i.e.\ second-order noise).         \label{tab1}}
\end{table*}

\section{Summary and Conclusions\label{sec:conclusions}}
In summary, with the transmon we have proposed a new type of superconducting qubit: the transmission-line shunted plasma oscillation qubit. In terms of the value of the energy ratio $E_J/E_C\sim10^2$, the transmon is intermediate between a CPB qubit and a current-biased phase qubit, but unlike either of these has no dc connections. At the same time, it may be viewed as a natural improvement of the CPB qubit, which provides its underlying quantum circuit.

The realization of the vision of quantum computing depends crucially on the design of physical systems which satisfy the DiVincenzo criteria \cite{divincenzo}, in particular the requirements of sufficiently long coherence time and scalability. Remarkably, the transmon should offer significant improvements with respect to both of these requirements: (i) we predict a strong improvement of insensitivity with respect to charge noise in comparison to the CPB, see Table \ref{tab1}; (ii) the drastically improved charge insensitivity should make locking to the charge degeneracy point unnecessary, thus simplifying the setup of multi-qubit systems significantly. As we have shown, the key to the favorable properties of the transmon lies in the combination of exponential decrease of the charge dispersion, the slow power-law decay of the anharmonicity, and the realization of strong coupling to the transmission line resonator. In addition, the dispersive regime of the coupled system is described by an ac-Stark shift Hamiltonian in complete analogy to the regular CPB, allowing for the transfer of control and readout protocols from the CPB to the transmon system.

The effort to reduce the noise susceptibility in solid-state qubits based on Josephson junctions has lead to a variety of different qubit types. Usually, these designs achieve a noise suppression in one particular channel, oftentimes accompanied by a tradeoff with respect to noise in other channels. Flux qubits \cite{friedman,wal} operate at $E_J/E_C$ ratios similar to those of the transmon, i.e.\ $E_J/E_C\approx10^2-10^3$. Accordingly, flux qubits reach an insensitivity to charge noise comparable to the transmon. However, flux qubits will typically show a significantly larger susceptibility to flux noise, especially when operated away from the flux sweet spot. Phase qubits \cite{martinis} trade in a slight increase in critical-current noise sensitivity for a drastic suppression of charge noise. Recent devices using inductive coupling to establish a current bias \cite{steffen} may also face increased flux sensitivity. 

Remarkably, the transmon achieves its exponential insensitivity to $1/f$ charge noise without incurring a penalty in the form of increased sensitivity to either flux or critical-current noise. This advantage can be illustrated by comparing the transmon to the traditional CPB, as shown in Table \ref{tab1}. 
As discussed above, the transmon is in fact comparatively less sensitive to flux and critical-current noise than the CPB. In fact, even without any reduction in the canonical $1/f$ noise intensities, we predict that a transmon qubit operated at the flux sweet spot should be limited only by the effects of relaxation. 
In conclusion, we are confident that the transmon will belong to a new generation of superconducting qubits with significantly improved coherence times and scalability.

\begin{acknowledgments}
We would like to thank Robert Willett for useful input on surface acoustic waves, and Patrice Bertet, Hans Mooij, John Martinis, Luigi Frunzio, and Blake Johnson for valuable discussions.  This work was supported in part by Yale University via a Quantum Information and Mesoscopic Physics Fellowship (JK and AAH), by the Department of Defense (TMY), by NSERC, FQRNT and CIAR (AB), by NSA under ARO contract number W911NF-05-1-0365, and the NSF under grants ITR-0325580 and DMR-0603369.
\end{acknowledgments}

\appendix
\section{Full network analysis\label{app:network}}
\begin{figure}
    \centering
  \includegraphics[width=0.9\columnwidth]{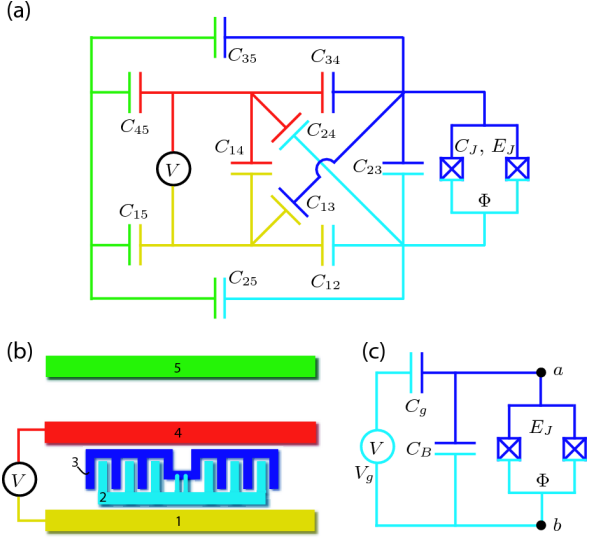}
    \caption{(Color online) (a) Full capacitance network for the transmon device. 
    (b) Simplified schematic of the transmon device design (not to scale). (c) Reduced network. \label{fig:network}}
\end{figure}

For completeness, we describe the modeling of the transmon device starting from an analysis of the full capacitance network \cite{footnote4}. This network is depicted in Fig.~\ref{fig:network}(a). It is based on the capacitances between the various conducting islands, see Fig.~\ref{fig:network}(b). For a minimal model, we take into account the two ground planes and center pin of the transmission line resonator as well as the two islands connected through the Josephson junctions. In the actual device, the dc bias is supplied via an additional capacitance to the center pin. For simplicity, we restrict our network to five islands in Fig.~\ref{fig:network}, considering only the effective voltage $V$ between bottom ground plane and center pin. 

By Th\'{e}venin's theorem, any single-port linear network of impedances and voltage sources can be substituted by an equivalent circuit consisting of one voltage source $V'$ and one impedance. In our particular case it is useful to retain the original gate voltage source $V_g$ in the equivalent circuit. This can be accomplished by allowing for one additional impedance, as shown in Fig.~\ref{fig:network}(c). The two effective capacitances can be interpreted as an effective gate capacitance $C_g$ and an effective shunting capacitance $C_B$. Together, they adjust for the correct voltage seen from the Josephson-junction port via the parameter $\beta=V_{ab}/V_g=C_g/C_\Sigma$ and the total capacitance $C_\Sigma=C_B+C_g+C_J$ between the nodes $a$ and $b$; see Fig.~\ref{fig:network}(c). [In the following, we absorb the junction capacitance into $C_B$.]

The parameters $\beta$ and $C_\Sigma$ are extracted from the full capacitance network as follows. Each conducting island, enumerated by $i=1,\ldots,n$, is associated with a certain charge $Q_i$ and a potential $\phi_i$ (with respect to infinity). These obey the linear relation $Q_i=\sum_j C_{ij}\phi_j$. For each island, we know either its charge or its potential. Let us choose the island enumeration such that for islands $i\le i_0$, the charges $Q_i^*$ are known, whereas for $i>i_0$ the potentials $\phi_i^*$ are known. (Here, the additional star signals that the quantity is known.) We thus have the following system of linear equations:
\begin{align}
Q_i^*&=\sum_{j\le i_0} C_{ij}\phi_j + \sum_{j>i_0} C_{ij}\phi_j^*\quad \text{for } i\le i_0\\
Q_i&=\sum_{j\le i_0} C_{ij}\phi_j + \sum_{j>i_0} C_{ij}\phi_j^*\quad \text{for } i> i_0,
\end{align}
from which we can determine the unknown quantities $\phi_1,\ldots,\phi_{i_0},Q_{i_0+1},\ldots,Q_n$. With the solution, we can immediately calculate the voltage exhibited at the $a-b$ port by subtracting the corresponding island potentials. This yields the splitting parameter $\beta$. The total capacitance $C_\Sigma$ is obtained from the full network by substituting the voltage source by a short and calculating the total charging energy of the network when applying a voltage $V_J$ across the junction. Equating the result with $C_\Sigma V^2_J/2$, one obtains the parameter $C_\Sigma$.

The treatment of the transmon embedded in a transmission line resonator is only slightly more complicated. Again, the use of Th\'{e}venin's theorem allows for the reduction of the capacitance network to a few effective capacitances, see Fig.~\ref{fig:circuit-diagram}(a). Here, the effect of the resonator can be modeled by a local LC oscillator \cite{blais1}. Following the standard quantization procedure for circuits \cite{devoret2}, we obtain
\begin{align}\label{ham2}
\hat{H}&=\frac{\hat{\phi}_r^2}{2L_r} + \frac{(C_B+C_g)\hat{Q}_r^2}{2C_*^2}\\\nonumber
&+\frac{(C_g+C_{in}+C_r)\hat{Q}_J^2}{2C_*^2}-E_J\cos\left(\frac{2\pi}{\hbar} \hat{\phi}_J\right)\\\nonumber
&+\frac{C_g\hat{Q}_r\hat{Q}_J}{C_*^2}+\frac{(C_BC_{in}+C_gC_{in})\hat{Q}_rV_g + C_gC_{in}\hat{Q}_JV_g}{C_*^2}
\end{align}
For simplicity, we have absorbed the junction capacitances into the parallel capacitance $C_B$, and introduced the abbreviation 
\[
C_*^2=C_BC_g + C_BC_{in}+ C_gC_{in}+C_BC_r+C_gC_r.
\]
In Eq.~\eqref{ham2} the first two terms describe the local oscillator of the resonator, the two terms in the second line capture the qubit's degrees of freedom and the terms in line 3 give the coupling between the two of them and the coupling to the gate electrode. Taking into account that $\hat{V}=V^0_\text{rms}(\hat{a}+\hat{a}^\dag)$ and assuming that $C_r\gg C_B,\,C_{in},\,C_g$, we recover the Hamiltonian \eqref{ham1}.

\section{Mathieu solution for the CPB Hamiltonian\label{app:mathieu}}
We briefly review the solution of the Hamiltonian \eqref{CPB-gen} in terms of Mathieu functions, generalizing the results from Refs.~\cite{cottet, devoret1} to arbitrary values of the effective offset charge $n_g$. In the phase basis, the stationary Schr{\"o}dinger equation is given by
\be\label{phase_ham}
\left[ 4E_C\left(-i\frac{d}{d\varphi}-n_g\right)^2 -E_J\cos \varphi\right]\psi(\varphi)=E\psi(\varphi),
\ee
where the boundary condition is $\psi(\varphi)=\psi(\varphi+2\pi)$. We can recast Schr{\"o}dinger's equation in the standard form of Mathieu's equation by introducing the function $g(x)\equiv e^{-2in_g x}\psi(2x)$, so that
\be\label{math-eq}
g''(x) + \left[\frac{E}{E_C}+\frac{E_J}{E_C}\cos(2x)\right]g(x)=0.
\ee
The $2\pi$-periodicity of $\psi(\varphi)$ translates into a pseudo-periodicity of $g(x)$ with characteristic exponent $\nu=-2(n_g-k)$, where $k\in\ZZ$. Following the notation of Meixner and Sch\"afke \cite{meixner2}, Eq.~\eqref{math-eq} is solved by the Floquet-type solution $\me_\nu(q=-\frac{E_J}{2E_C},x)$. Accordingly, the eigenenergies $E$ are fixed by Mathieu's characteristic value, see Eq.~\eqref{energies}, and the wavefunctions can be represented as
\be
\psi_m(\varphi)=\frac{\exp (in_g \varphi)}{\sqrt{2}}\me_{-2[n_g-k(m,n_g)]}\left(-\frac{E_J}{2E_C},\frac{\varphi}{2}\right).
\ee

The integer numbers $k$ have to be chosen in such a way to correctly sort the eigenenergies and eigenstates. This implies that $k$ becomes a function of the band index $m$ and the effective offset charge $n_g$. Extending Cottet's treatment \cite{cottet} to cover the full range $n_g\in\RR$, we find that this function is given by
\begin{align}
k(m,n_g)&=\sum_{\ell=\pm1} [\rnd(2n_g+\ell/2) \mmod 2]\\\nonumber
&\qquad\times[\rnd(n_g)+\ell(-1)^m\{ (m+1) \divg 2\}].
\end{align}
Here, $\rnd(x)$ rounds to the integer closest to $x$, $a\mmod b$ denotes the usual modulo operation, and $a \divg b$ gives the integer quotient of $a$ and $b$.

\section{Perturbation theory for the large $E_J/E_C$ limit\label{app:perturb}}
For completeness, we briefly review the perturbative approach employed in Sections \ref{sec:anharmonicity} and \ref{sec:cqed} for large $E_J/E_C$. Starting from the Hamiltonian in the phase basis, Eq.~\eqref{phase_ham}, one notes that the Josephson energy acts as a strong ``gravitational force'' on the rotor, effectively restricting the angle $\varphi$ to small values around zero. This motivates (i) the neglect of the periodic boundary condition, and (ii) the expansion of the cosine for small angles. Keeping terms up to fourth order, this yields the potential energy $-E_J+E_J\varphi^2/2-E_J\varphi^4/24$.

The Hamiltonian can now be viewed as a harmonic oscillator with a quartic perturbation describing the leading-order anharmonicity. Due to (i), the ``vector potential'' $n_g$ can be eliminated by a gauge transformation, and the resulting Hamiltonian can be cast in the form of a Duffing oscillator 
\be
H=\sqrt{8E_CE_J}(\hat{b}^\dag \hat{b} +1/2) - E_J -\frac{E_C}{12}(\hat{b}+\hat{b}^\dag)^4,
\ee
where $\hat{b}$, $\hat{b}^\dag$ denote the regular annihilation and creation operators for the harmonic oscillator approximating the transmon. The leading-order correction to the eigenenergies arising from the quartic term is given by
\begin{align}
E_j^{(1)} &= -\frac{E_C}{12}\boket{j}{(\hat{b}+\hat{b}^\dag)^4}{j}%\\\nonumber
           = -\frac{E_C}{12} (6j^2+6j+3).
\end{align}
Note that in this section, $\ket{j}$ denotes the pure harmonic oscillator state in the absence of any anharmonicity.
The leading-order correction to the state $\ket{j}$,
\be
\ket{j}^{(1)} = -\frac{E_C}{12}\sum_{i\not=j} \frac{\boket{i}{(\hat{b}+\hat{b}^\dag)^4}{j}}{E_i-E_j}\ket{i},
\ee
causes a mixing of $\ket{j}$ with the states $\ket{j\pm4}$ and $\ket{j\pm2}$.

Noting that for large $E_J/E_C$ the Cooper pair number operator can be reexpressed in terms of 
\be
\hat{n}=-i\left(\frac{E_J}{8E_C}\right)^{1/4}\frac{1}{\sqrt{2}}(\hat{b}-\hat{b}^\dag),
\ee
it is straightforward to evaluate the asymptotic expressions for the matrix elements of $\hat{n}$, $\hat{n}^2$ involved in the charge fluctuations (see Section \ref{sec:numberphase}) and the coupling strength (see Section \ref{sec:coupl}).

\section{Canonical transformation\label{app1}}
The elimination of the interaction term in Eq.~\eqref{hamiltonian-before} to lowest order in $g_{i,i+1}/\Delta_i$ (in the dispersive limit) can be accomplished by a canonical transformation $\hat{H}'=\hat{D}\hat{H}\hat{D}^\dag$. Writing the unitary operator in the form $\hat{D}=\exp[\hat{S}-\hat{S}^\dag]$, we find that the appropriate generator for the lowest three qubit states is given by
\be
\hat{S}=\sum_i \beta_i \hat{a} \ket{i+1}\bra{i}.
\ee
Here,  the parameters $\beta_i$ are defined as
\begin{align}
\beta_i &=\frac{g_{i,i+1}}{\omega_{i,i+1}-\omega_r}=\frac{g_{i,i+1}}{\Delta_i}.
\end{align}
Employing the Baker-Campbell-Hausdorff relation and retaining terms up to order $g_{ij}^2/\Delta_i^2$, we obtain the transformed Hamiltonian
\begin{align}\label{effectiveham}\nonumber
\hat{H}'=& \sum_i\hbar\omega_i\ket{i}\bra{i} + \hbar\omega_r \hat{a}^\dag \hat{a} 
+ \sum_i \hbar\chi_{i,i+1}\ket{i+1}\bra{i+1}\\\nonumber
&-\hbar\chi_{01}\hat{a}^\dag \hat{a} \ket{0}\bra{0} + \sum_{i=1}^\infty \hbar(\chi_{i-1,i}-\chi_{i,i+1})\hat{a}^\dag \hat{a} \ket{i}\bra{i}\\
&+\sum_i\hbar \eta_i\hat{a}\hat{a}\ket{i+2}\bra{i} +\text{h.c.}
\end{align}
with
\be
\eta_i=\frac{g_{i,i+1}g_{i,i+2}[(\omega_{i+1}-\omega_{i+2})-(\omega_{i}-\omega_{i+1})]}{2(\omega_{i+1}-\omega_i-\omega_r)(\omega_{i+2}-\omega_{i+1}-\omega_r)}
\ee
and $\chi_{ij}$ defined in Eq.~\eqref{chieq}. The terms in the last line of Eq.~\eqref{effectiveham} describe two-photon transitions, and are negligible as compared to the remaining terms due to the smallness of the parameters $\eta_i$.


\begin{thebibliography}{48}
\bibitem{footnote1}
\bibnamefont{It is interesting to note that the $\alpha=0$ point (``anharmonicity barrier'') smoothly varies with $n_g$ and vanishes below/above a certain $n_g$ threshold.}
\bibitem{footnote3}
\bibnamefont{We note that results for the anharmonicity have been previously presented in Ref.\ \cite{cottet}.}
\bibitem{footnotenew}
\bibnamefont{We emphasize that this mechanical analogy is only meant to explain the fundamental difference between dc and ac response, not the enhancement of the coupling strength when increasing $E_J/E_C$.}
\bibitem{footnote4}
\bibnamefont{Based on simple estimates, we conclude that the effect of residual inductances is negligible, so that the capacitance network is a sensible starting point.}
\end{thebibliography}
\end{document}